\documentclass[12pt]{article}
\usepackage{amssymb,amsmath,bm,mathrsfs,makeidx,amsfonts,graphicx,amsthm,color}
\usepackage[round]{natbib}
\usepackage{epsfig}
\usepackage{rotating}
\usepackage{mathtools}
\usepackage{bm}
\usepackage{extarrows}
\usepackage[colorlinks,linkcolor=red,anchorcolor=blue,citecolor=blue]{hyperref}
\usepackage{booktabs}
\usepackage{xr}
\usepackage{chngpage}

\numberwithin{equation}{section}
\def\beginn{\begin{eqnarray*}}
\def\endn{\end{eqnarray*}}
\def\beginy{\begin{eqnarray}}
\def\endy{\end{eqnarray}}
\def\begine{\begin{enumerate}}
\def\ende{\end{enumerate}}

\def\be{\begin{equation}}
\def\ee{\end{equation}}
\def\bea{\begin{eqnarray}}
\def\eea{\end{eqnarray}}

\numberwithin{equation}{section}
\theoremstyle{plain}
\newtheorem{thm}{Theorem}[section]
\newtheorem{prop}{Proposition}[section]
\newtheorem{lem}{Lemma}[section]
\newtheorem{coro}{Corollary}[section]
\newtheorem{rmk}{Remark}[section]

\newtheorem{assu}{Assumption}[section]

\newcommand{\non}{\nonumber \\}
\newcommand{\bbA}{{\bf A}}

\newcommand{\bbB}{{\bf B}}

\newcommand{\bbw}{{\bf w}}
\newcommand{\bbI}{{\bf I}}

\newcommand{\bbT}{{\bf T}}

\newcommand{\bbX}{{\bf X}}

\newcommand{\bbx}{{\bf x}}

\newcommand{\bbY}{{\bf Y}}
\newcommand{\bby}{{\bf y}}
\newcommand{\bbZ}{{\bf Z}}

\newcommand{\bbmu}{{\boldsymbol\mu}}

\newcommand{\E}{\mathbb{E}}
\newcommand{\R}{\mathbb{R}}
\newcommand{\vv}{\mathrm{vech}}
\newcommand{\pr}{\mathbb{P}}

\newcommand{\var}{\mathrm{Var}}

\begin{document}

\title{Estimating a change point in a sequence of very high-dimensional covariance matrices}
 \author{H. Dette, G. M. Pan and Q. Yang}

\date{}
\maketitle

\begin{abstract}
This paper considers the problem of estimating a change point in the covariance matrix in a sequence of high-dimensional vectors, where the dimension is substantially larger than the sample size.

A two-stage approach is proposed to efficiently estimate the location of the change point. The first step consists of a   reduction of the  dimension to identify elements of the covariance matrices corresponding to significant changes. In a second step we use   the components after dimension reduction to determine the position of  the change point. Theoretical properties are developed for both steps and numerical studies are conducted to support the new methodology.
\end{abstract}

\vspace{10pt}
\textbf{Keywords}: High-dimensional covariance matrices; change point analysis; dimension reduction.

\section{Introduction} \label{sec1}

Change point detection has a long history having it origins in quality control [see \cite{wald1945} or \cite{page1954,page1955} for early references] and
it has been an active field of research until today since the phenomena of sudden changes arise in various areas, such as financial data
(house market, stock), signal processing, genetic engineering, seismology, machine learning.
 In the last decades numerous authors have worked on this problem from several perspectives including the construction of tests for   the hypotheses of the existence of change points
and the estimation of their locations. We refer to \cite{auehor2013} and \cite{jandhyala:2013} for some recent reviews on this subject.

An important  problem in the detection of structural breaks in multivariate data
 is the detection of changes in a sequence of  means.  \cite{chu1996}, \cite{horkokste1999}, \cite{horvhusk2012} and \cite{kirmusomb2015} investigated  this problem   using different
variants of CUSUM statistics. More recently, the   high-dimensional case
 (dimension larger than the sample size) has been discussed by several authors as well. \cite{jirak2015} considered a maximum of statistics across panels coordinate-wise
 to test the hypothesis of at least one change point in a sequence of high dimensional mean vectors
 [see also \cite{detgoes2017} who studied   relevant changes in this context]. \cite{chofryz2015}  suggested sparsified binary segmentation for this problem, while \cite{Cho2016}
 investigated a double CUSUM approach  transferring - roughly speaking -  the high-dimensional data to a univariate CUSUM statistic.
 We also mention the work of \cite{enikharch2014}, who looked at the problem under sparse alternatives and   \cite{wangsam2018}, who
 considered the situation, where  at  certain  time  points,  the
mean structure changes in a sparse subset of the coordinates.

While substantial effort has been spent on change point analysis for the multivariate mean, the problem of detecting structural breaks
in the covariance matrix  has not been studied so intensively in the literature. For a fixed dimension, say $p$,  \cite{auehor2013} developed  nonparametric change point analysis based on the well known CUSUM approach.
 \cite{Dette2016} proposed a general approach to detect relevant change points in a parameter of a time series. In an online supplement to this paper a test for a relevant change in the covariance matrix is proposed, where the dimension is also fixed.
Recently, \cite{kaotrapurga2018} considered the case where the dimension is increasing with the sample size and
demonstrated by means of a simulation study that tests of stability of the whole covariance matrix have severe size distortions. As an alternative they proposed and investigated change point analysis
based on PCA. In an unpublished preprint  \cite{avanbuzu2016}  also looked at the high-dimensional setting and suggested  a multiscale approach
under sparsity assumptions, while \cite{wangyurin2017} considered the problem of detecting multiple change points in the situation $p=O(n/\log n) $ (here $n$ is the sample size) and
investigated optimality properties of the   binary segmentation  [see \cite{vostrikova1981}] and the wild binary segmentation  algorithm [see \cite{Fryzlewicz14}] for localising  multiple changes points in a sequence
of high dimensional covariance matrices.

The purpose of the present  paper is to propose an alternative  estimator of the change point  in  a sequence of  very high dimensional  covariance matrices and to investigate its theoretical and empirical properties,  where we do not impose any sparsity assumptions on the matrices.
 To be precise, suppose that
 $\bbx_1,\cdots,\bbx_{k_0}, \bbx_{k_0+1},\cdots,\bbx_n$
 are $p$-dimensional   observations with common mean vector $\bbmu$ and existing covariance matrices.
 The parameter $k_0$ defines  the true change point in the structure of the covariance matrices. That is,
the first $k_0$ observations have covariance matrix $\Sigma_1=(\sigma_1{(a,b)})_{a,b=1,\ldots,p} \in \R^{p\times p}$,
while the last $n-k_0$ observations have covariance matrices $\Sigma_2=(\sigma_2({a,b}))_{a,b=1,\ldots,p} \in \R^{p\times p}$ and $\Sigma_1\neq \Sigma_2$. We are interested in estimating the point $k_0$. One difficulty in dealing with changes in the covariance matrix is the dimensionality since there are $p(p+1)/2$ positions needed to be compared.  This brings in much noisy information when there are many equal components in the matrices $\Sigma_1$ and $\Sigma_2$, leading to a loss of accuracy in detection of the change point. Taking this consideration into account, we propose to proceed in two steps to identify the change point.
First, we  apply a dimension reduction technique reducing the dimension from $p(p+1)/2$ in the original problem to a substantially
 smaller value, say $m$.   Roughly speaking, we   only keep the components
 in the analysis for which a weighted mean of the squared differences of the covariance estimators from the samples
 $\bbx_1,\cdots,\bbx_{k}$ and  $\bbx_{k+1},\cdots,\bbx_n$ exceeds a given threshold (the mean is calculated summing with respect to  the different values of the potential change points of $k$).
 Therefore our approach is vaguely related to the estimation of sparse covariance matrices, which has found considerable attention in the literature
 [see \cite{bicklevi2008b},  \cite{lamfan2009}  or \cite{fanliaoliu2016} among many others]. However, in contrast to this work, we do not assume a sparse structure of the covariance matrix, but identify important components by  thresholding  a weighted sum of the (squared) differences corresponding to all potential samples before and after a postulated change point.
In a second step after dimension reduction, we use a CUSUM type statistic based on the reduced components to  locate the change point.

An outline of the paper is given as follows. In Section \ref{sec2}, we introduce our main methodology -- both the dimension reduction step and the detection step.
In particular, a  bootstrap method is suggested to select the threshold used for the dimension reduction  (see the discussion in Section \ref{boot}).
Theoretical results are developed in Section \ref{sec3}, where we prove that (asymptotically) we identify all relevant components correctly and that we estimate the location of the change $k_0$ consistently.     In Section \ref{sec4} we investigate the finite sample properties of the new method and demonstrate that it yields precise estimates of the change point in situations, where the dimension is substantially larger than the sample size. We also provide a comparison with two alternative methods which are most similar in spirit to our approach and have recently been proposed in the literature.
Finally, all proofs and technical details are deferred to an appendix in Section  \ref{sec5}.

\section{Methodology}\label{sec2}
Let  $\bbx_1,\cdots,\bbx_{k_0},\bbx_{k_0+1},\cdots,\bbx_n$ denote a sample of independent $p$-dimensional random vectors   with common mean $\bbmu$ and existing covariance matrices. The position $k_0$ is the ``true'' change point of the covariance matrices, i.e.
the first $k_0$ random variables $\bbx_1,\cdots,\bbx_{k_0}$  have covariance matrix $\Sigma_1=(\sigma_1{(a,b)})_{a,b=1,\ldots,p} \in \R^{p\times p}$,
while the last $(n-k_0)$ random variables $ \bbx_{k_0+1},\cdots,\bbx_n$  have covariance matrices $\Sigma_2=(\sigma_2({a,b}))_{a,b=1,\ldots,p} \in \R^{p\times p}$ and $\Sigma_1\neq \Sigma_2$.  Our aim is to estimate  the location $k_0$ of the change.
For this purpose  we proceed in two steps.
\begin{itemize}
\item {\bf Step 1} consists of a dimension reduction. If
$$\hat \Sigma^k_1 = (\hat \sigma^{k}_1 ({a,b} ) )^p_{a,b=1}$$
   and $\hat{\Sigma}^n_{k+1} = ( \hat \sigma_{k+1}^n  ({a,b}) )^p_{a,b=1}$ denote the respective estimators of the covariance matrices from the data $\bbx_1,\ldots,\bbx_k$ and $\bbx_{k+1},\ldots, \bbx_n$,
    we - roughly speaking - only keep components in the change point analysis for which the quantity
 $$
 \sum^{n-2}_{k=2} k(n-k) \Big( \hat \sigma^{k}_1({a,b})  - \hat \sigma^n_{k+1} ({a,b}) \Big)^2
 $$
 is sufficiently large.
\item {\bf Step 2} consists of the  detection of a change point in the data obtained after dimension reduction. For this purpose let $\tilde \sigma^k_1$ and $\tilde \sigma^n_{k+1}$ denote the vectors  containing all  elements of the matrices  $\hat\Sigma^k_{1}$ and $\hat \Sigma^n_{k+1}$ corresponding to components  which have been identified in the first step of the procedure. Then - roughly speaking - we propose to estimate the change point by maximizing the statistic
    $$
    \tilde U_n(k) = \big \| (n-k)k (\tilde \sigma^k_1 - \tilde \sigma^n_{k+1} ) \big \|^2_2
    $$
    where $\| \cdot \|_2$ denotes the Euclidean norm.
\end{itemize}
We will give a detailed explanation of these two steps in the following subsections, where the statistics under consideration will be slightly modified.
The proposed methodology depends on a regularisation parameter, say $\tau$,  determining the amount of dimension reduction for Step 1,  and in Section \ref{boot} we introduce a data-driven
method for choosing this  threshold.

\subsection{Dimension reduction  }\label{step1}
For $i=1,\ldots,n$ denote  by $\bbx_i=(X_{i1},\cdots,X_{ip})^T$ the $i$th observation, let $\bar\bbx=\frac{1}{n}\sum\limits_{i=1}^n \bbx_i=(\bar X_1,\cdots,\bar X_p)$ be the sample mean and define
\begin{eqnarray}\label{yq10.1}
\dot{\bbx}_i=\bbx_i-\bar\bbx ~=~\big ( \dot{X}_{i1}, \ldots , \dot{X}_{ip} \big )^T ~=  \big  (X_{i1}-\bar X_1 , \ldots  , X_{ip}-\bar X_p \big )^T
\end{eqnarray}
as the vector of centered observations. We introduce the following  statistic
\begin{eqnarray}\label{yq1-2}
V_k=  \big ( V_k (a,b) \big ) _{1 \leq a \leq b \leq p} &=&
\frac{1}{k(k-1)}\mathop{\sum\sum}_{i\neq j\leq k}\vv(\dot{\bbx}_i\dot{\bbx}_i^T)\circ \vv(\dot{\bbx}_j\dot{\bbx}_j^T) \\
&&+\frac{1}{(n-k)(n-k-1)}\mathop{\sum\sum}_{i\neq j>k}\vv(\dot{\bbx}_i\dot{\bbx}_i^T)\circ \vv(\dot{\bbx}_j\dot{\bbx}_j^T)\non
&&-\frac{2}{k(n-k)}\sum_{i\leq k}\sum_{j>k}\vv(\dot{\bbx}_i\dot{\bbx}_i^T)\circ \vv(\dot{\bbx}_j\dot{\bbx}_j^T),\nonumber
\end{eqnarray}
{ where ``vech($H$)'' indicates the half-vectorization $p(p+1)/2$ vector by vectorizing only the lower triangular part of the symmetric matrix $H=(H(a,b))^p_{a,b=1}$ } and ``$\bbx\circ\bby$'' is the Hadamard product (or entrywise product) of the vectors $\bbx$ and $\bby$. Obviously,
 $V_k$ is a $p(p+1)/2$-dimensional vector.

We first give an intuitive illustration of the motivation behind the construction of the statistic $V_k$ defined in \eqref{yq1-2}, which
is in fact  motivated by being an  approximation of the statistic
\begin{eqnarray*}
\tilde V_k&=&\frac{1}{k^2}\mathop{\sum}_{i, j=1}^k\vv(\dot{\bbx}_i\dot{\bbx}_i^T)\circ \vv(\dot{\bbx}_j\dot{\bbx}_j^T)
+\frac{1}{(n-k)^2}\mathop{\sum}_{i, j=k+1}^n\vv(\dot{\bbx}_i\dot{\bbx}_i^T)\circ \vv(\dot{\bbx}_j\dot{\bbx}_j^T) \non
&&-\frac{2}{k(n-k)}\sum_{i=1}^k\sum_{j=k+1}^n\vv(\dot{\bbx}_i\dot{\bbx}_i^T)\circ \vv(\dot{\bbx}_j\dot{\bbx}_j^T) \\
&=&  \Big ( \frac{1}{k}\mathop{\sum}_{i=1}^k\vv(\dot{\bbx}_i\dot{\bbx}_i^T)  - \frac{1}{n-k}\mathop{\sum}_{i=k+1}^n\vv(\dot{\bbx}_i\dot{\bbx}_i^T) \Big ) \circ \Big ( \frac{1}{k}\mathop{\sum}_{i=1}^k\vv(\dot{\bbx}_i\dot{\bbx}_i^T)  - \frac{1}{n-k}\mathop{\sum}_{i=k+1}^n\vv(\dot{\bbx}_i\dot{\bbx}_i^T) \Big ). \nonumber
\end{eqnarray*}
Note that   the vector $\tilde V_k$ coincides with the vector
$$
\big( \big ( \hat \sigma^{k}_1({a,b}) - \hat \sigma_{k+1}^n({a,b}) \big )^2    \big)_{1 \leq  a \leq b \leq p}
$$
of the squared  (componentwise) differences of the elements of the covariance estimators
$\hat \Sigma^k_1 = \frac {1}{k} \sum^k_{i=1} \dot{\bbx}_i \dot{\bbx}^T_i$ and
$\hat \Sigma^n_{k+1} = \frac {1}{n-k} \sum^n_{i=k+1} \dot{\bbx}_i \dot{\bbx}^T_i$. Consequently,
at the ``true''  change point   $k=k_0$, one can verify that $\tilde V_{k_0}$ is an
estimator of vech$(\Sigma_1-\Sigma_2)^2$, which will be used to measure the difference between the two population covariance matrices.
The difference between  $\tilde V_k$ and $V_k$ consists  in the fact that  in the statistic $V_k$ we omit the terms corresponding to $i=j$ to eliminate the influence of  the
covariances of the random variables $\vv(\dot{\bbx}_i\dot{\bbx}_i^T)$.

However, in the change point problem we actually do not know the location of $k_0$, and we have to consider all the positions $k$
as long as the statistic is well defined. In particular, we obtain for the expectation of the  component $V_k(a,b)$ of the vector $V_k$
    corresponding to the   position $(a,b)$ in the matrices $\Sigma_1$ and $\Sigma_2$
\begin{eqnarray*}\label{yq1-5}
\E V_k(a,b) = \left\{
  \begin{array}{l l}
    (1-\frac{2}{n})^2(\sigma_1({a,b})-\sigma_2({a,b}))^2 & \mbox{ if } k=k_0,\\
    \frac{k_0(k_0-1)}{k(k-1)}(1-\frac{2}{n})^2(\sigma_1({a,b})-\sigma_2({a,b}) )^2 & \mbox{ if } k>k_0,\\
    \frac{(n-k_0)(n-k_0-1)}{(n-k)(n-k-1)}(1-\frac{2}{n})^2(\sigma_1({a,b})-\sigma_2({a,b}) )^2 & \mbox{ if } k<k_0.
  \end{array}
\right.
\end{eqnarray*}
 Note also, that the
  mean of $V_k(a,b)$ always achieves the largest value $$(1-\frac{2}{n})^2(\sigma_1({a,b})-\sigma_2({a,b}) )^2$$
  at the true change point position $k_0$ because the coefficients $ \frac{k_0(k_0-1)}{k(k-1)}$ and $\frac{(n-k_0)(n-k_0-1)}{(n-k)(n-k-1)}$ before $(1-\frac{2}{n})^2(\sigma_1({a,b})-\sigma_2({a,b}))^2$ are smaller than 1  when $k\neq k_0$. Moreover, these coefficients  are only related to
$k$ and are not influenced by the position $(a,b)$. Consequently, for any fixed $k$, larger values of $V_k(a,b)$ indicate a larger difference between
$\sigma_1{(a,b)}$ and $\sigma_2({a,b})$, thus implying a significant component.
Additionally, instead of investigating each value of $k$ separately, we suggest a weighted sum
\begin{equation}\label{yq1-3}
D=\frac{1}{n-3}\sum_{k=2}^{n-2}\frac{k(n-k)}{n}V_k,
\end{equation}
  to identify   the largest   components among  the $p(p+1)/2$ entries.
 The weights $\frac{k(n-k)}{n}$ are introduced to address the different sizes of   the variance of $V_k$ for different values of $k$. By selecting the largest entries in the vector $D$, we are able to identify the components with the largest changes.

In view of this discussion, we conduct the dimension reduction as follows.
Let $D(a,b)$ denote the elements of the vector $D$ corresponding to the position $(a,b)$ in the matrices $\Sigma_1$ and $\Sigma_2$.
We determine   all  components which are larger than a critical value $\tau$, which will be   specified in Section \ref{boot},   define
\begin{equation}\label{yq1-4}
{\cal D}_\tau=\{(a,b): D(a,b) > \tau, 1\leq a\leq b \leq p\}
\end{equation}
as the set of all corresponding components and denote by $ m=\# {\cal D}_\tau$  its cardinality.
In this way, we reduce the $p(p+1)/2$-dimensional vector to
a vector of   dimension $m$.    In the next step  we will simply work with the $m$-dimensional vectors corresponding to the components  identified by the set ${\cal D}_\tau$. In Theorems \ref{theorem1} and \ref{theorem1a}, it will be shown that after dimension reduction with an appropriate
threshold all  entries with no difference are discarded, while all the entries with a sufficiently large difference   are kept.

\subsection{Change point detection after dimension reduction}
For the estimation of the change point we propose the test statistic
{
\begin{equation}\label{yq1-7}
U_n(k)=\frac{1}{n^4}\mathop{\sum^{k}\sum^{k}}_{(i\neq t)=1}\mathop{\sum^{n}\sum^{n}}_{(j\neq l)=k+1}
(\dot{\widetilde{\bbx}}_i-\dot{\widetilde{\bbx}}_j)^T(\dot{\widetilde{\bbx}}_t-\dot{\widetilde{\bbx}}_l),
\end{equation}
}
where $\dot{\widetilde{\bbx}}_i$ is an $m$-dimensional subvector of $\vv(\dot{\bbx}_i\dot{\bbx}_i^T)$, only keeping the $m$ components of the index set
${\cal D}_\tau$ defined in (\ref{yq1-4}). Then the estimator of the  change point $k_0$ is defined by
\begin{equation}\label{yq1-8}
\hat{k}=\underset{1\leq k\leq n}{\arg\max}\ \ U_n(k).
\end{equation}
The motivation behind the construction of $U_n(k)$ stems from the fact the statistic $U_n(k)$ is related to a CUSUM type
statistic which is frequently used in change point analysis. To be precise, consider the CUSUM statistic
\begin{eqnarray*}
\tilde U_n(k) &=& \frac {1}{n^4} \sum^k_{i,t=1} \sum^{n}_{j,l=k+1} (\dot{\widetilde{\bbx}}_i - \dot{\widetilde{\bbx}}_j)^T (\dot{\widetilde{\bbx}}_t - \dot{\widetilde{\bbx}}_l) \\
&=& \frac {1}{n^4} \Big \{ (n-k)^2 \Big(\sum^k_{i=1} \dot{\widetilde{\bbx}}_i \Big)^T \Big(\sum^k_{j=1} \dot{\widetilde{\bbx}}_j \Big) - 2k(n-k) \Big(\sum^k_{i=1} \dot{\widetilde{\bbx}}_i \Big)^T \Big(\sum^n_{j=k+1} \dot{\widetilde{\bbx}}_j \Big) \\
& \qquad & + k^2 \Big( \sum^n_{i=k+1} \dot{\widetilde{\bbx}}_i \Big)^T \Big( \sum^n_{j=k+1} \dot{\widetilde{\bbx}}_j \Big) \Big \} \\
&=& \frac {1}{n^4} \Big  \| (n-k) \sum^k_{i=1} \dot{\widetilde{\bbx}}_i - k \sum^n_{j=k+1} \dot{\widetilde{\bbx}}_j  \Big \|^2_2
~=~ \frac {1}{n^4}
\big \| (n-k)k (\tilde \sigma^k_1 - \tilde \sigma^n_{k+1} ) \big \|^2_2
\end{eqnarray*}
where
$\| \cdot \|_2$ denotes the usual Euclidean norm   and
$\tilde \sigma^k_1 $  and $  \tilde \sigma^n_{k+1} $ respectively denote the vectors containing the elements of the covariance estimators
 $\hat \Sigma^k_{1} $   and $ \hat \Sigma_{k+1}^n$ corresponding to positions identified in the first step.
 Observing the definition of $\dot{\widetilde{\bbx}}_i$ and noting that the difference
\begin{eqnarray*}
\tilde U_n(k) - U_n(k) &=& \frac {1}{n^4} \sum^k_{i=1} \sum^n_{j=k+1} \| \dot{\widetilde{\bbx}}_i - \dot{\widetilde{\bbx}}_j \|^2_2 +
 \frac {1}{n^4} \sum^k_{i=1} \sum^{n}_{(j\neq l)=k+1} ( \dot{\widetilde{\bbx}}_i - \dot{\widetilde{\bbx}}_j )^T  ( \dot{\widetilde{\bbx}}_i - \dot{\widetilde{\bbx}}_l ) \\
  & \qquad & + \frac {1}{n^4}   \sum^{k}_{(i\neq t)=1} \sum^n_{j=k+1}( \dot{\widetilde{\bbx}}_i - \dot{\widetilde{\bbx}}_j )^T  ( \dot{\widetilde{\bbx}}_t - \dot{\widetilde{\bbx}}_j )
\end{eqnarray*}
{is of smaller order than $U_n(k)$ when $k$ is far from $1$ and $n$}, we see that the statistic $U_n(k)$ is a CUSUM type statistic obtained from the components identified in the first step. It is therefore related to the statistic in equation (2.12) in  \cite{aueetal2009}, who proposed an estimator of the change point based on a quadratic form using {\bf ALL} elements of the difference $\hat \Sigma^k_{1} - \hat \Sigma_{k+1}^n$.
Note that in the definition of $U_n$ we eliminate the influence of the covariances $cov(\dot{\widetilde{\bbx}}^T_i\dot{\widetilde{\bbx}}_i)$ by omitting terms corresponding to $i=t$ and $j=l$ in $\tilde U_n(k)$. As a consequence, we avoid the estimation of such   higher order moments.

We will show in Theorem \ref{theorm2} that - under appropriate regularity conditions -  the statistic
$\hat k$ in \eqref{yq1-8} is in fact a consistent estimator   of the change point  if the dimension and sample size converge to infinity
and the threshold is chosen appropriately. More precisely,  we can choose $\tau=C\cdot\max(\log p,\log n)$ with a sufficiently large constant $C$
and  the dimension $p$ can be of polynomial order of the sample size $n$.

\subsection{Selecting the threshold $\tau$ via resampling }\label{boot}
For a data driven choice of the threshold  $\tau$ we propose  a bootstrap   approach, which mimics the distributional properties in the case of no
change point.
 To be precise define
\begin{eqnarray}\label{yq1-6}
&&\bbT= \big (\bbT_1,\cdots, \bbT_n \big)= \big (\vv(\dot\bbx_1\dot\bbx_1^T),\vv(\dot\bbx_2\dot\bbx_2^T),\cdots,\vv(\dot\bbx_n\dot\bbx_n^T) \big)
\in \R^{\frac{p(p+1)}{2}\times n},\non
&&
\bbZ= {\frac{1}{\sqrt 2}}
\big (\bbT_{2}-\bbT_{1},  \bbT_{4}-\bbT_{3}, \ldots , \bbT_{2\lfloor \frac{n}{2} \rfloor } -\bbT_{2\lfloor \frac{n}{2} \rfloor-1} \big )  \in \R^{\frac{p(p+1)}{2}\times \lfloor \frac{n}{2} \rfloor },
\end{eqnarray}
and denote by  $\bbZ_{(i)}^T$  the $i$-th row of the matrix $\bbZ$,  i.e.
$$
\bbZ=  \big (\bbZ_{(1)},\cdots, \bbZ_{(p(p+1)/2)} \big)^T.
$$
For each $\ell =1,\cdots, p(p+1)/2$, the $\lfloor \frac{n}{2} \rfloor $-dimensional vector $\bbZ_{(\ell )}$ can be considered as a combination of $\lfloor \frac{n}{2} \rfloor $
 observations and   we  denote  the empirical  standard deviation of these  $\lfloor \frac{n}{2} \rfloor $ observations by $o_{\ell \ell }$.
  The construction of the matrix $\bbZ$ ensures  that  the means of its columns are zero  except for at most one position
  (note that the means of the columns of $\bbT$ have one change point at $k_0$), which does not have a substantial effect on the standard deviation
 provided that the sample size is not too small.
If the variance of the random variables in the $\ell$th  row of the matrix $\bbT$ is constant, it is easy to see that $o_{\ell \ell }^2$ estimates
this variance. If this assumption is not satisfied, generally speaking, $o_{\ell \ell }^2$ always estimates $n\times \var(\bar T_{(l)})$, where $\bar T_{(l)}$ is the average
of the random variables in the $\ell$th  row of the matrix $\bbT$. Note that
 the  factor $\frac{1}{\sqrt{2}}$  in \eqref{yq1-6} reflects the fact that the matrix $\bbZ$ is formed from differences of two consecutive columns of
 the matrix $\bbT$.

In order to  estimate  the threshold $\tau$   let $\Upsilon $ denote  the
 $\frac{p(p+1)}{2}\times \frac{p(p+1)}{2}$  diagonal matrix, with entries $o_{11}, \ldots , o_{p(p+1)/2, p(p+1)/2}$.
We generate a new data matrix $\bbY=(y_{ij}) \in \R^{p(p+1)/2 \times n}$  with independent  standard normal distributed entries and define
$$\bbX^*=\Upsilon\bbY.$$
In other words, in the bootstrap we replace  the quantities $\vv(\dot\bbx_1\dot\bbx_1^T) , \ldots , \vv(\dot\bbx_n\dot\bbx_n^T) $
 by the $n$ columns of $\bbX^{\ast}$. As a consequence  the terms  $\dot X_{1a}\dot X_{1b} , \ldots , \dot X_{na}\dot X_{nb}$
are replaced by  $o_{\ell \ell }y_{\ell 1},\ldots, o_{\ell \ell  } y_{\ell n}$ where the random variables $y_{\ell1},\ldots, y_{\ell n}$
are independent standard normal distributed and the index $\ell  $ corresponds to the position $(a,b)$.

Next we calculate for the matrix $\bbX^*$  the quantities $V^*_k$ and $D^*$ defined in (\ref{yq1-2}) and (\ref{yq1-3}) respectively,  and obtain the vector
\[
D^*=(D^* (a,b) : 1 \leq a \leq b \leq p).
\]
The  threshold  $\tau$ is finally defined  as the largest entry of $D^*$, i.e.
\[
\tau=\max_{1 \leq a \leq b \leq p} D^*(a,b).
\]

\section{Asymptotic properties} \label{sec3}

In this section we discuss the theoretical properties of our approach. For this purpose we need several assumptions, which will be stated first, beginning with conditions on the distribution of the random vectors $\bbx_i$.

\begin{assu}\label{cond1} {\rm
Denote $\bbx_i=(X_{i1},\cdots,X_{ip})$, $i=1,\cdots,n$. For any $1\leq a\leq p$, $X_{ia}$ is a  sub-Gaussian random variable, i.e.
there are positive constants $C_1, C_2$ (independent of the indices $i$ and $a$) such that for every $t>0$,
\[
\pr\left(|X_{ia}|>t\right)\leq C_1e^{-C_2t^2}.
\]
Moreover, the covariance matrices before and after the change point
satisfy $\|\Sigma_{\nu}\|_{op}\leq M$  ($\nu=1,2$) for some  positive constant $M$, {where $\|\cdot\|_{op}$ denotes the spectral norm.}
}
\end{assu}

\noindent
Our next assumption specifies the size of the change, which can be detected using the threshold $\tau$. Note that the dimension $p$ is increasing with the sample size
and a difference between the matrices $\Sigma_1$ and $\Sigma_2$ might vanish asymptotically if $p,n \to \infty $
although it is visible for any  fixed $p$ (for example if $\Sigma_1 - \Sigma_2 = e_pe^T_p/p$ where $e^T_p = (0,\ldots, 0,1)$).

\begin{assu}\label{cond4}
{\rm The smallest nonzero entry of the matrix $\Sigma_1-\Sigma_2$ satisfies}
\begin{equation}\label{yq0410.3}
| \sigma_1({a,b})-\sigma_2({a,b}) | >
\lambda\geq C\sqrt{\frac{\tau}{n}}\max\Big\{\frac{n^2}{(n-k_0)^2},\ \frac{n^2}{k_0^2}\Big\}.
\end{equation}
\end{assu}

\noindent
Note that condition (\ref{yq0410.3})
implies that
\begin{equation}\label{yq0531.1}
\lambda\geq C\sqrt{\frac{\tau}{n}}\max\Big\{\frac{n}{n-k_0},\ \frac{n}{k_0},\ \sqrt{\frac{n}{k_0}}, \ \sqrt{\frac{n}{n-k_0}},\ \frac{n}{k_0}\sqrt{\frac{n-k_0}{k_0}},\
\frac{n}{n-k_0}\sqrt{\frac{k_0}{n-k_0}}\Big\}.
\end{equation}

\begin{assu}\label{cond5}
{\rm For some small positive constant $c$ we have}
\begin{eqnarray}
p^2n&=&o(e^{c\tau }) \label{c1} \\
p^2n&=&o(e^{cn^{\frac{1}{4}}\sqrt{\tau} }) \label{c2} \\
 p^2n^2 &=&o(e^{c\sqrt{\tau} }). \label{c3}
\end{eqnarray}
\end{assu}

\begin{thm}\label{theorem1}
Recall the definition of the set $\mathcal{D}_\tau$ in \eqref{yq1-4} and define
$$
{\cal N} = \{(a,b) \colon 1 \leq a \leq b \leq p; \ \sigma_1({a,b}) =\sigma_2({a,b}) \}
$$
as the set of indices corresponding to equal elements in the matrices $\Sigma_1$ and $\Sigma_2$. Then    under Assumption \ref{cond1}
\begin{equation}\label{yq001exp}
\pr\Big \{\bigcup_{(a,b)\in {\cal N}} \{D(a,b)>\tau\}\Big\} = \pr \big ( {\cal N} \cap {\cal D}_\tau \not = \emptyset \big )
\leq c_1p^2n\left[e^{-c_2\tau}+e^{-c_2n^{1/4}\sqrt{\tau }}+ne^{-c_2\sqrt{\tau}}\right],
\end{equation}
where $c_1$ and $c_2$ are some constants. In particular, if Assumption    \ref{cond5} is also satsified
\begin{equation}\label{yq001}
\pr\Big \{\bigcup_{(a,b)\in {\cal N}} \{D(a,b)>\tau\}\Big\}
 \rightarrow 0,
\end{equation}
i.e, after dimension reduction, all the entries with no difference are discarded.
\end{thm}

\begin{thm}\label{theorem1a}
Recall the definition of the set $\mathcal{D}_\tau$ in \eqref{yq1-4} and define
$$
{\cal P} = \{ (a,b) : 1 \leq a \leq b \leq p; \mid\sigma_1({a,b})-\sigma_2({a,b})\mid > \lambda \}
$$
as the set of components which differ by more than $\lambda$. Then under Assumption \ref{cond1} and   \ref{cond4} we have
\begin{equation}\label{yq002exp}
\pr\Big\{\bigcap_{(a,b) \in {\cal P}} \{D(a,b)>\tau\}\Big\} = \mathbb{P} (\mathcal{P} \subset \mathcal{D}_\tau) \geq
1-c_3p^2n\left[e^{-c_4\tau}+e^{-c_4n^{1/4}\sqrt{\tau }}+ne^{-c_4\sqrt{\tau}}\right],
\end{equation}
where $c_3$ and $c_4$ are  constants. In particular, if Assumption    \ref{cond5} is also satisfied
\begin{equation}\label{yq002}
\pr\Big\{\bigcap_{(a,b) \in {\cal P}} \{D(a,b)>\tau\}\Big\}
\rightarrow 1,
\end{equation}
i.e, after dimension reduction, all components corresponding to a difference larger than $\lambda$ are kept.
\end{thm}

 In our next result we establish the asymptotic consistency of the estimator $\hat k$. Here and throughout this paper the symbol $\stackrel{i.p.}\longrightarrow$ denotes convergence in probability.

\begin{thm}\label{theorm2}
If Assumptions \ref{cond1} and   \ref{cond4} are satisfied, we have
$$
\pr\Big\{\Big|\tfrac{\hat k}{k_0}-1\Big|\geq\epsilon\Big\}\leq
c_5p^2n\left[e^{-c_6\tau}+ne^{-c_6\sqrt{\tau}}\right],
$$
where $c_5$ and $c_6$ are  constants.
 In particular, if Assumption    \ref{cond5} is also satisfied it follows
$$
\frac{\hat k}{k_0}\stackrel{i.p.}\longrightarrow 1.
$$
\end{thm}

\begin{coro}\label{coro1} Theorem \ref{theorem1}, Theorem \ref{theorem1a} and Theorem \ref{theorm2} are still true if
Assumption \ref{cond5} is replaced by the following Assumption \ref{condnew}.
\end{coro}
\begin{assu}\label{condnew}
{\rm Assume that $k_0>n^{\epsilon}$ for some $0<\epsilon<1$ and that there exists a positive constant $M<\infty$ such that
$pn^{-M}\rightarrow 0$ and  $pn e^{-c\tau }\rightarrow 0$ for some sufficiently small positive constant $c$. Note that $M$ could be any large positive constant.}
\end{assu}

\begin{rmk}
{\rm
 Note that we can choose $\tau=C\cdot\max(\log p,\log n)$ in  Assumption \ref{condnew}, where
$C$ is a sufficiently large constant. Then the only requirements are  $k_0>n^{\epsilon}$ for the location of the change and
 that the dimension $p$ cannot exceed a polynomial  order of the sample size $n$ (but the degree of the polynomial can be  arbitrary).
 Moreover, the inequality \eqref{yq0410.3}  also qualitatively describes a relation  between the location of the change and the
size of the differences between the elements of the covariance matrices before and after the change.
For example, if  $p>n$,  we have $\tau=C\cdot\max(\log p,\log n)=C\cdot \log p$  and, if
 $k_0$ is proportional to $n$, this means that the smallest non  zero element of the matrix $\Sigma_1 - \Sigma_2$ should satisfy
$$
 \mid \sigma_1({a,b})-\sigma_2({a,b}) \mid \  >  C\sqrt{\frac{\log p}{n}}.
$$
 This is a well-known order to  distinguish signal from noise in  covariance matrix estimation, see for example \cite{bicklevi2008b},
 who considered covariance estimators based on thresholding.
\\
On the other hand, the choice $\tau=C\cdot\max(\log p,\log n)$ is not  possible in Assumption \ref{cond5}.  However,
if we choose $\tau=C\cdot\max((\log p)^2,(\log n)^2)$ with a sufficiently large constant $C$, there is no restriction
on the dimension $p$ and $n$.
 Now, if $k_0$ is proportional to $n$, this means that the smallest non  zero element of the matrix $\Sigma_1 - \Sigma_2$ has  to satisfy
$$
 \mid \sigma_1({a,b})-\sigma_2({a,b})  \mid \  >  C \frac {\max(\log p,\log n)}{\sqrt{n}}
$$
for some constant $C$.  This means that the procedure estimates $k_0$ consistently  even if the differences between
the elements of the two matrices are very small.
}
 \end{rmk}

\section{Finite sample properties}\label{sec4}
In this section we investigate the finite sample properties of the new change point estimator by means of a simulation study
and compare our approach with two alternative methods proposed by \cite{aueetal2009} and \cite{avanbuzu2016}, which
are most similar in spirit as the procedure proposed in the present paper.

To be precise let $r_0=k_0/n$ be the ``true''  change point fraction and  let  $\hat r=\hat k/n$, where  $\hat k$
is the new change point estimator  defined   in \eqref{yq1-8}. All the numerical results below are calculated from $200$
replications  and we  obtain the simulated estimates  $\hat r_1,\ldots, \hat r_K$  of $r_0$.
In the following discussion we present
the mean
$$
\bar r=\frac{1}{K}\sum\limits_{i=1}^K \hat r_i~,
$$
the standard deviation
\[
std (\hat r) =\sqrt{\frac{1}{K-1}\sum_{i=1}^K (\hat r_i-\bar  r)^2}
\]
and the corresponding mean squared error
\[
MSE=\frac{1}{K}\sum_{i=1}^K (\hat r_i-r_0)^2 =\frac{K-1}{K}\cdot std^2  (\hat r) +(\bar r-r_0)^2.
\]
Throughout this section we denote  by  {\bf blk}$(\bbA,\bbB)$
 a block-diagonal matrix composed by matrices $\bbA$ and $\bbB$ of appropriate dimension.
 $\Sigma_1=\bbI_p$ is always  the identity matrix and we consider  four different choices for the matrix
 $\Sigma_2$ to investigate the performance of the new estimator under the following  alternatives
\begin{itemize}
\item {\bf case 1:} $\Sigma_2=1.5\ast\bbI_p$; \qquad \qquad \quad\ {\bf case 2:} $\Sigma_2=2\ast\bbI_p$;
\item {\bf case 3:} $\Sigma_2=\text{\bf blk}(4,\bbI_{p-1})$;\qquad\qquad  {\bf case 4:} $\Sigma_2=\text{\bf blk}(8,\bbI_{p-1})$.
\end{itemize}
Cases 1 and 2 indicate that there are many ($p$ positions) small disturbances between $\Sigma_1$ and $\Sigma_2$, with a
 magnitude increasing  from $0.5$ to $1$. On the other hand
  there is only one disturbance between the two population covariance matrices in cases 3 and 4, but the magnitude is more significant (3 and 7 respectively).

The true change point fraction is  chosen as  $r_0=k_0/n=0.5$ and  the first $k_0$ and  the last $(n-k_0)$   samples
are generated from a  multivariate normal distribution ${\cal N} _p(0, \Sigma_1)$ and  a  ${\cal N}_p(0, \Sigma_2)$ distribution, respectively.

\subsection{Performance of the new  estimator }\label{sec41}

In order to investigate the finite sample properties of the new estimator  we choose two sample sizes
$n=100$ and $n=200$ and consider different  dimensions $p$ ranging from $5$ to $500$. For each  pair $(n,p)$, the mean change point fraction,
standard deviation (std) and mean squared error (MSE) are recorded for all four cases under consideration,
and the results are summarized in Table \ref{n100} ($n=100$) and Table \ref{n200} ($n=200$).
 The numerical results from the two tables can be summarized as follows:

 \begin{itemize}
\item[(1)]
When the sample size $n$ increases, the  performance of the estimator  is better.
\item[(2)] The dimension $p$ of the data  does not have a significant influence on the performance of the estimators. In particular
the mean squared error is remarkably stable with respect to the dimension in all four cases under consideration.
\item[(3)]   When the magnitude of the disturbance between $\Sigma_1$ and $\Sigma_2$ increases,
 the estimator  performs   better (compare the results from  case 1 with case 2 or from case 3 with  case 4).
\end{itemize}

\begin{table}[!t]
\begin{center}
\caption{\label{n100}  \small
\it Mean, standard deviation (std) and mean squared error (MSE)  of the estimator
$\hat r = \hat  k / n$ defined in \eqref{yq1-8}. The sample size is  $n=100$,
the change point is $k_0=50$,  $\Sigma_1 = I_p$ and results of  four different choices  for   $\Sigma_2$
are presented.}
\medskip

{\scriptsize
\begin{tabular}{c|c c c c c c|| c c c c c c}
\hline
&\multicolumn{6}{c||}{{\bf case 1}}&\multicolumn{6}{c}{{\bf case 2}}\\
\hline
$p$&5&20&60&200&300&500&5&20&60&200&300&500\\
\hline
mean  &0.4985&0.5081& 0.5101  &  0.5213&0.5183 & 0.4830 & 0.5186 &   0.5169& 0.5059&  0.5017& 0.5080  & 0.5099\\
std &0.1669 &0.1475& 0.1274 & 0.1267 & 0.1246  & 0.1766 & 0.1032 & 0.0673& 0.0780&  0.0913 & 0.0667 & 0.0797\\
MSE &0.0277&    0.0217   & 0.0162 &   0.0164  &  0.0158  &  0.0313   & 0.0109 &   0.0048&    0.0061   & 0.0083 &   0.0045 &   0.0064\\
\hline
\hline
&\multicolumn{6}{c||}{{\bf case 3}}&\multicolumn{6}{c}{{\bf case 4}}\\
\hline
$p$&5&20&60&200&300&500&5&20&60&200&300&500\\
\hline
mean &0.5423 &0.5416&0.5393& 0.5275& 0.5378& 0.5410& 0.5345 &0.5288 &0.5271 &0.5325 &0.5317 &   0.5312\\
std &0.0570& 0.0654 &0.0649 &0.0491 & 0.0569& 0.0580& 0.0471 & 0.0364& 0.0527 & 0.0389 & 0.0416& 0.0457\\
MSE &0.0050   & 0.0060  &  0.0057  &  0.0031 &   0.0046&    0.0050 &   0.0034 &   0.0022 &   0.0035  &  0.0026 &   0.0027  &  0.0030\\
\hline
\end{tabular}
}
\end{center}
%
%
%
\begin{center}
\caption{\label{n200} \small
\it Mean, standard deviation (std) and mean squared error (MSE)  of the estimator
$\hat r = \hat  k / n$ defined in \eqref{yq1-8}. The sample size is  $n=200$,
the change point is $k_0=100$,   $\Sigma_1 = I_p$ and results of  four different choices  for   $\Sigma_2$
are presented.}

\medskip

{\scriptsize
\begin{tabular}{c|c c c c c c|| c c c c c c}
\hline
&\multicolumn{6}{c||}{{\bf case 1}}&\multicolumn{6}{c}{{\bf case 2}}\\
\hline
$p$&5&20&60&200&300&500&5&20&60&200&300&500\\
\hline
mean &0.5108 & 0.5064&  0.5071& 0.5058&0.5067 &0.5042&0.5105&0.5052 &0.5029&0.5010&0.5019&0.5010\\
std &0.1096&0.0625&0.0437 &0.0575 &0.0457&0.0529&0.0249&0.0128&0.0120&0.0032&0.0103& 0.0043\\
MSE &0.0121 &   0.0039 &   0.0020 &   0.0033   & 0.0021   & 0.0028 &   0.0007    &0.0002   & 0.0002 &   0.0000 &   0.0001  &  0.0000
\\
\hline
\hline
&\multicolumn{6}{c||}{{\bf case 3}}&\multicolumn{6}{c}{{\bf case 4}}\\
\hline
$p$&5&20&60&200&300&500&5&20&60&200&300&500\\
\hline
mean &0.5227&0.5234&0.5253 &0.5238& 0.5211&0.5227& 0.5192& 0.5175 &0.5156& 0.5203&0.5212 &0.5180\\
std &0.0320&0.0381 &0.0418&0.0350&0.0378& 0.0364 & 0.0307 &   0.0242  &  0.0235  &  0.0255 &   0.0288  &  0.0262\\
MSE &0.0015  &  0.0020 &   0.0024 &   0.0018  &  0.0019&    0.0018  &  0.0013 &   0.0009  &  0.0008  &  0.0011 &   0.0013  &  0.0010\\
\hline
\end{tabular}
}
\end{center}
\end{table}

Next we investigate the influence of   the dimension reduction step on performance
of  the estimator. To this end, we consider  case 1 and case 3 with sample sizes  $n=200$
and present   in Table \ref{noreduc} the corresponding results without dimension reduction.
In other words we apply the estimator (\ref{yq1-8}) based on all components.
We note that the computation time without dimension reduction  is substantially larger
because we work with $p(p+1)/2$-dimensional vectors.

Comparing Table \ref{noreduc} with the corresponding  results  in Table \ref{n200}, we observe the
following.
 \begin{itemize}
\item[(1)]  In case 1, the  differences in the bias of  $\hat r$ are  negligible (in both cases the mean is very close to $0.5$).
 On the other hand the standard deviations and as consequence the MSE in Table \ref{noreduc}
are smaller, which means that the   estimator without dimension reduction is more stable.
Note that there are many small disturbances between two population matrices and thus keeping all positions
promotes a safer estimation.
On the other hand  the MSE results in Table \ref{n200}  from the estimator using dimension reduction
are already satisfactory.
 \item[(2)] In  case 3 when there is only one significant disturbance the situation is different. Although
 the bias  of $\hat r$ in Table \ref{noreduc} is smaller, its
 standard deviation and MSE  increase very  fast to an unacceptable level with increasing dimension.
 This means that the estimator without dimension reduction is not reliable if the dimension is large.
 \item[(3)] As an interesting phenomenon we note  that in contrast to  Table \ref{n200}
 the standard deviation and mean squared error  in Table \ref{noreduc} show a downward trend in case 1
 but  an upward tendency in case 3. This observation can be explained by
 the fact that without dimension reduction the gap between $\Sigma_1$ and $\Sigma_2$ increases
 with the dimension $p$ in case 1 but decreases in case 3.
\end{itemize}

\begin{table}[!htp]
\begin{center}
\caption{\label{noreduc} \it  \small
Mean, standard deviation (std) and mean squared error (MSE)  of the estimator
$\hat r = \hat  k / n$ without dimension reduction (i.e.  $\hat k$ is defined in \eqref{yq1-8} with
${\cal D}_\tau = \{ (a,b) \colon  1 \leq a  \leq b  \leq p \})$. The sample size is  $n=200$,
$\Sigma_1 = I_p$ and results for
two different choices  for   $\Sigma_2$
are presented ({case 1} and {case 3}) .
 }

 \medskip
{\scriptsize
\begin{tabular}{c|c c c c c c|| c c c c c c}
\hline
&\multicolumn{6}{c||}{{\bf case 1}}&\multicolumn{6}{c}{{\bf case 3}}\\
\hline
$p$&5&20&60&200&300&500&5&20&60&200&300&500\\
\hline
mean &0.5226 &   0.5068  &  0.5028 &   0.5037  &  0.5052&    0.5014 &  0.5265 &   0.5234   & 0.5211  &  0.5097   & 0.5098 &   0.5024\\
std &0.0758 &   0.0262 &   0.0193 &   0.0139   & 0.0162  &  0.0101 &   0.0420 &   0.0361 &   0.0429   & 0.1114  &  0.1243    & 0.1614\\
MSE &0.0062 &   0.0007   & 0.0004  &  0.0002 &   0.0003   & 0.0001&    0.0025   & 0.0018 &   0.0023 &   0.0125&    0.0155 &   0.0259\\
\hline
\end{tabular}
}
\end{center}
\end{table}

\subsection{Comparison with an estimator based on a quadratic form}
\label{sec42}
\begin{figure}[!htp]
    \centering
    \includegraphics[width=7.5cm,height=4.8cm]{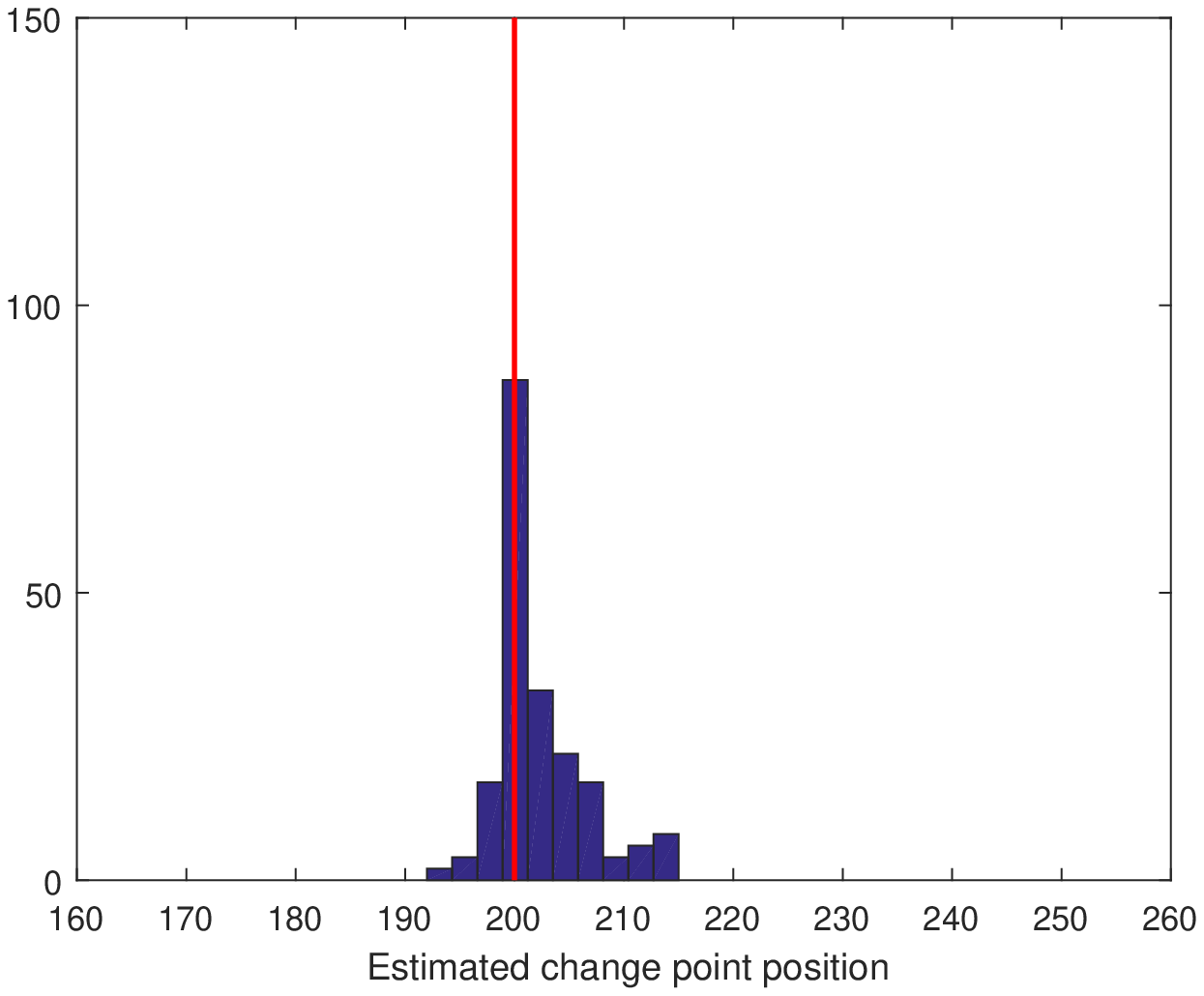}  ~~~  \includegraphics[width=7.5cm,height=4.8cm]{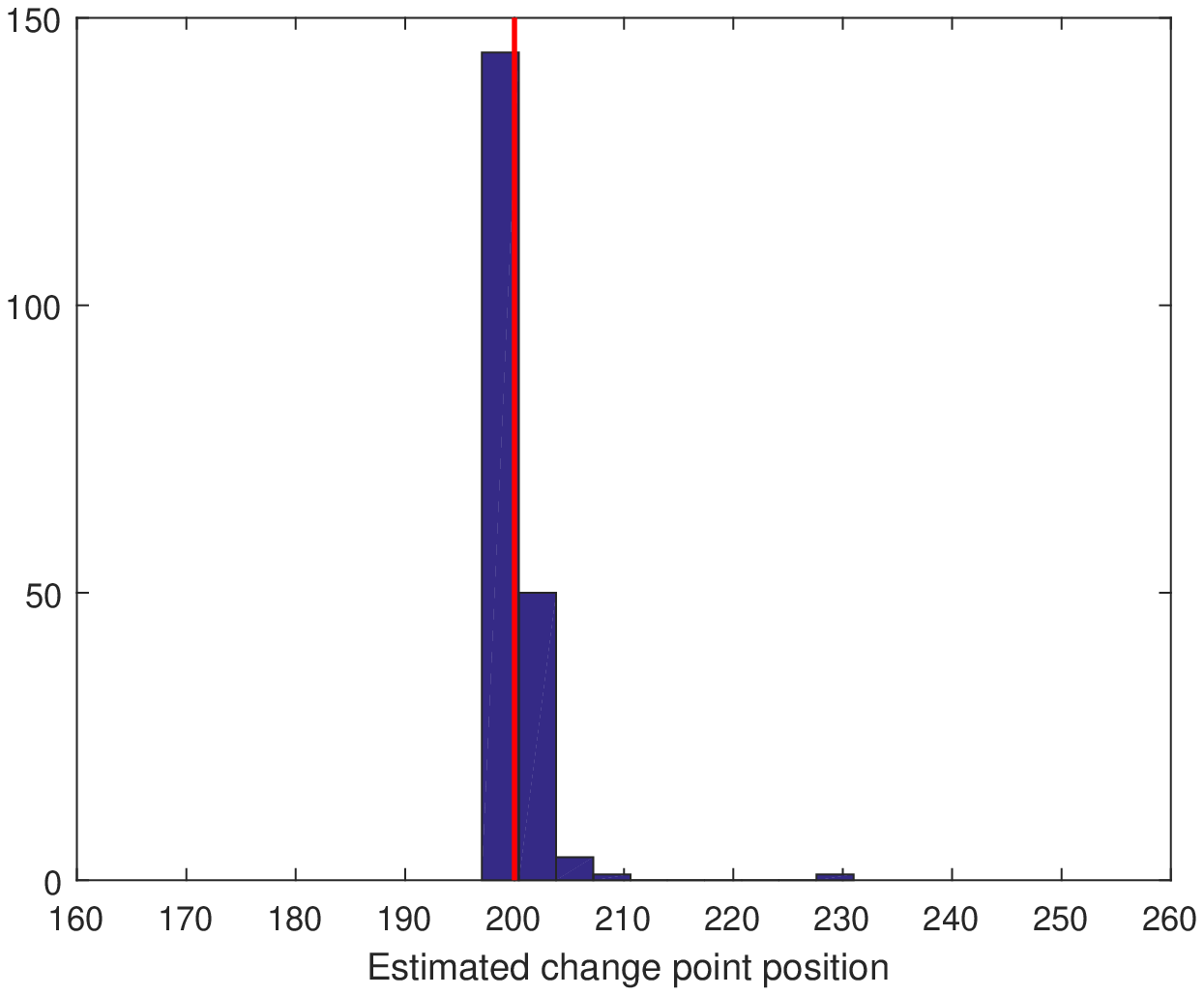}
    \includegraphics[width=7.5cm,height=4.8cm]{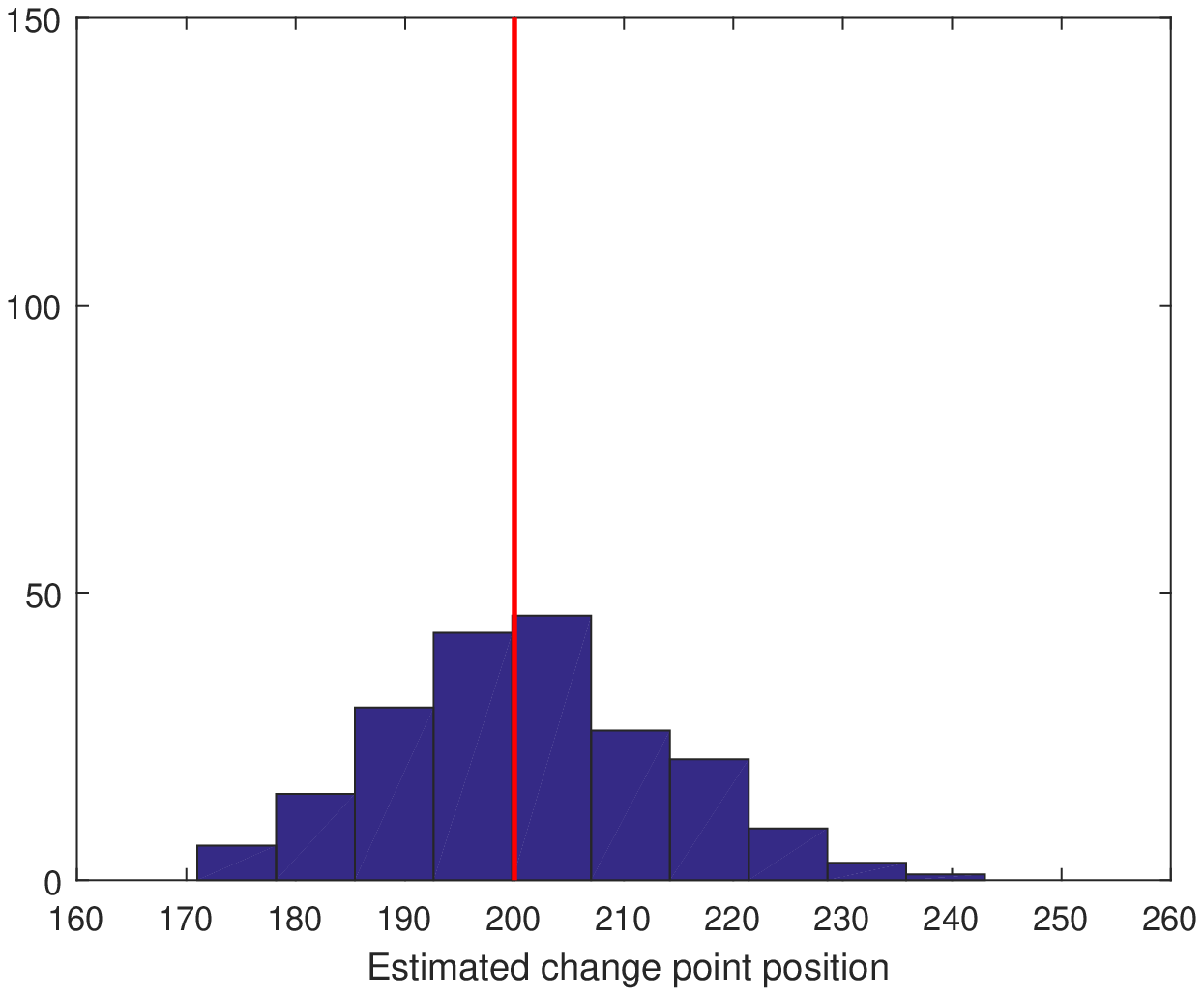}   ~~~ \ \includegraphics[width=7.5cm,height=4.8cm]{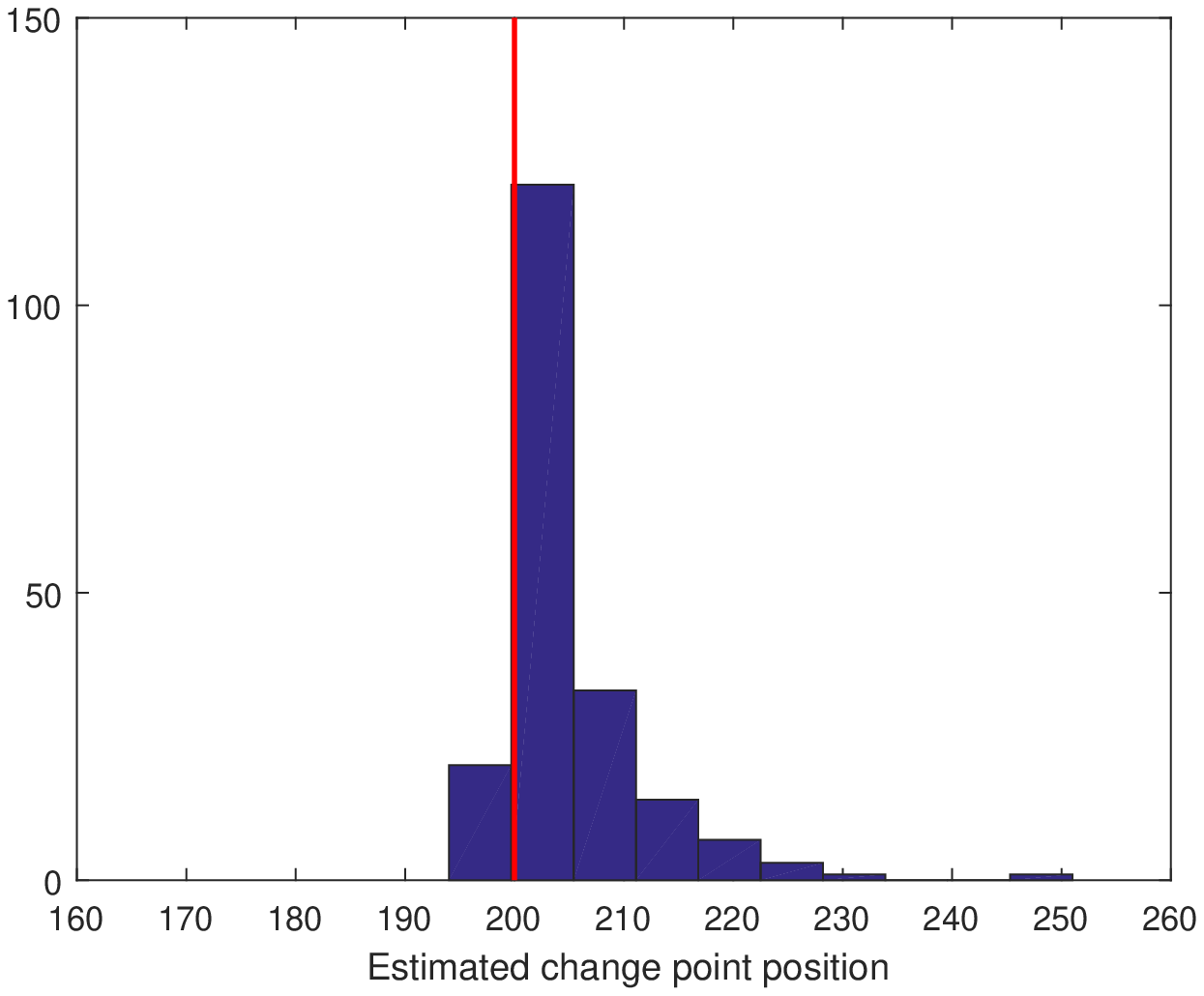}
    \caption{    \label{fig2} \small
 {   \it Histograms of estimated change point positions for the estimator of  \cite{aueetal2009} (left panels) and
 the new estimator \eqref{yq1-8}. Upper row: case 2, lower row: case 3.  The sample size is $n=400$, the
  dimension is  $p=25$ and the  change point is located at  $k_0=200$ (red line).}}
  \end{figure}

We first compare the new method with the estimator (2.12) suggested in \cite{aueetal2009}.
Note that this statistic involves an inverse matrix $\hat\Sigma_n^{-1}$, where $\hat\Sigma_n$ is an estimator of the long-run covariance satisfying their condition (2.6). Under our setting, this long-run covariance reduces to the population covariance matrix of a $p(p+1)/2$-dimensional random vector.
As consequence the dimension $p$ has to be substantially  smaller than the sample size $n$  to estimate the inverse
of the covariance matrix precisely. In order to get a larger range for the dimension  $p$, we use  the sample size  $n=400$ in this subsection and let
the dimension $p$ vary from $5$ to $50$ (for larger values of $p$ the method of \cite{aueetal2009} shows some instabilties).
The location of the change point is assumed to be $k_0=200$.
 In Table \ref{Auelarger} we display the results of the estimator  in \cite{aueetal2009} and the estimator  \eqref{yq1-8}  proposed  in this paper, where we restrict to the case
 2 and case  3 for the sake of brevity (the cases 1 and 4 show a very similar picture).
We observe that the new estimator always performs better. While this superiority is only minor for small dimension, it becomes substantial for $p=40, 50$, in this case
the mean squared error of the estimator  in  (2.12)  suggested by   \cite{aueetal2009} is  very large  (compared to the cases $p \leq 20$), while
the new estimate shows a remarkable stability with respect to different dimensions. The differences are also visualized  in Figure \ref{fig2}   for the
cases   2 and 3, respectively, where we show the histograms of both estimates obtained from the different simulation runs.
 The sample size is $n=400$, the  dimension is  $p=25$ and the  change point is located at  $k_0=200$ (red line).

\begin{table}[!htp]
\begin{center}
\caption{\label{Auelarger} \it
\small
Mean, standard deviation (std) and mean squared error (MSE)  of the estimator
(for the relative location $k_0/n$ of the change point)
proposed in  \cite{aueetal2009}  (left part) and the estimator $\hat k/n$ suggested in this paper (right part).
The sample size is  $n=400$, $\Sigma_1 = I_p$ and results for two  different choices  of
the matrix  $\Sigma_2$ are displayed.}

\medskip

{\footnotesize
\begin{tabular}{c|c c c c c|| c c c c c}
\hline
{\bf case 2}&\multicolumn{5}{c||}{ \cite{aueetal2009} }&\multicolumn{5}{c}{estimate \eqref{yq1-8} }\\
\hline
$p$&5&10&20&40&50&5&10&20&40&50\\
\hline
mean  &   0.5099 & 0.5043 & 0.5078   & 0.5800 &   0.6092 &  0.5048 &   0.5022 &   0.5012&    0.5005 &   0.5003\\
std &    0.0164 &   0.0103 &   0.0132 &   0.1092  &  0.1184   &  0.0116 &   0.0060  &  0.0037 &   0.0015&    0.0013\\
MSE &   0.0004   & 0.0001  &  0.0002 &   0.0183  &  0.0259  &    0.0002 &   0.0000  &  0.0000&    0.0000 &   0.0000\\
\hline
\hline
{\bf case 3}&\multicolumn{5}{c||}{ \cite{aueetal2009} }&\multicolumn{5}{c}{estimate \eqref{yq1-8} }\\
\hline
$p$&5&10&20&40&50&5&10&20&40&50\\
\hline
mean  &  0.5167  &  0.5148 &   0.5067 &   0.4988   & 0.4909 &  0.5145  &  0.5112 &   0.5128&    0.5135 &   0.5125\\
std &  0.0236 &   0.0351 &   0.0387  &  0.1386 &   0.1432  &    0.0270    &0.0173  &  0.0200&    0.0197   & 0.0209\\
MSE &    0.0008 &   0.0014 &   0.0015 &   0.0191 &   0.0205 &  0.0009   & 0.0004&   0.0006 &  0.0006 &   0.0006\\
\hline
\end{tabular}
}
\end{center}
\end{table}

\subsection{Comparison with a multiscale estimator } \label{sec43}

We conclude this section with a brief comparison with the procedure in \cite{avanbuzu2016} who proposed a multiscale
approach for the localization of the   change point.
For the sake of comparison  we use  the same design as in  Section 5.1 of this paper.  We also  performed a comparison under scenarios considered
in Section \ref{sec42}. Here the method proposed by these authors does not yield reliable estimates of the change point and
the  results are not displayed for the sake of brevity.
\begin{table}[!htp]
\begin{center}
\caption{\label{Avane2} \small
\it  Comparison of  the estimate  \eqref{yq1-8} with the change point estimate proposed by \cite{avanbuzu2016}.
Both estimates are normalized, that is  $\hat r=\hat\tau/n$,
where $\hat \tau $ is one of the two estimates. The sample size is   $n=1000$, the dimension is  $p=50$ and the ``true'' change point  is
located at $k_0/n =500/1000=0.5$.}
\medskip
\medskip
\begin{tabular}{ c c c|| c c c }
\hline
\multicolumn{3}{c||}{\cite{avanbuzu2016}}&\multicolumn{3}{c}{estimate \eqref{yq1-8}}\\
\hline
mean &std&MSE& mean &std&MSE\\
\hline
0.4878&0.0948&0.0091&0.5006&0.0087&0.0001\\
\hline
\end{tabular}
\end{center}
\end{table}

 To be precise  we summarize the setting  in \cite{avanbuzu2016}   here briefly.  The covariance matrix before  the change point is  $\Sigma_1=\bbI_p$
   and the  matrix  $\Sigma_2$  after the change point is  generated as follows.  First a Poisson distributed random variable $K\sim Poiss(3)$
   is generated. Then  the matrix $\Sigma_2$ is composed as a block-diagonal matrix
of $K$ (symmetric) matrices of size $2\times 2$ with ones on their diagonals and their off-diagonal element drawn uniformly from the set  $[-0.6;-0.3]\cup[0.3;0.6]$.
The remaining  $(p-2k)\times(p-2k)$ diagonal block  of  the  matrix  $\Sigma_2$ is the identity matrix and all other elements of $\Sigma_2$ are $0$.
We consider a  sample of   $n=1000$ observations with  dimension $p=50$, where the ``true''  change point
is given by   $k_0=n/2=500$. The procedure of  \cite{avanbuzu2016} also requires the specification of a  set  $\mathcal{I}_s$  corresponding to observations without change points
and we use $\mathcal{I}_s=[1,2,\cdots,100]$ (as suggested in their paper) in our simulation.  Note also that according to (2.5) in Section 2.3 of \cite{avanbuzu2016}
a change point estimator is  only defined if there exists a narrowest window detecting a change-point. This was in $199$ of  the $200$ replications the
 case.

\begin{figure}[!htp]
    \centering
     \includegraphics[height = 5.5cm, width=7cm]{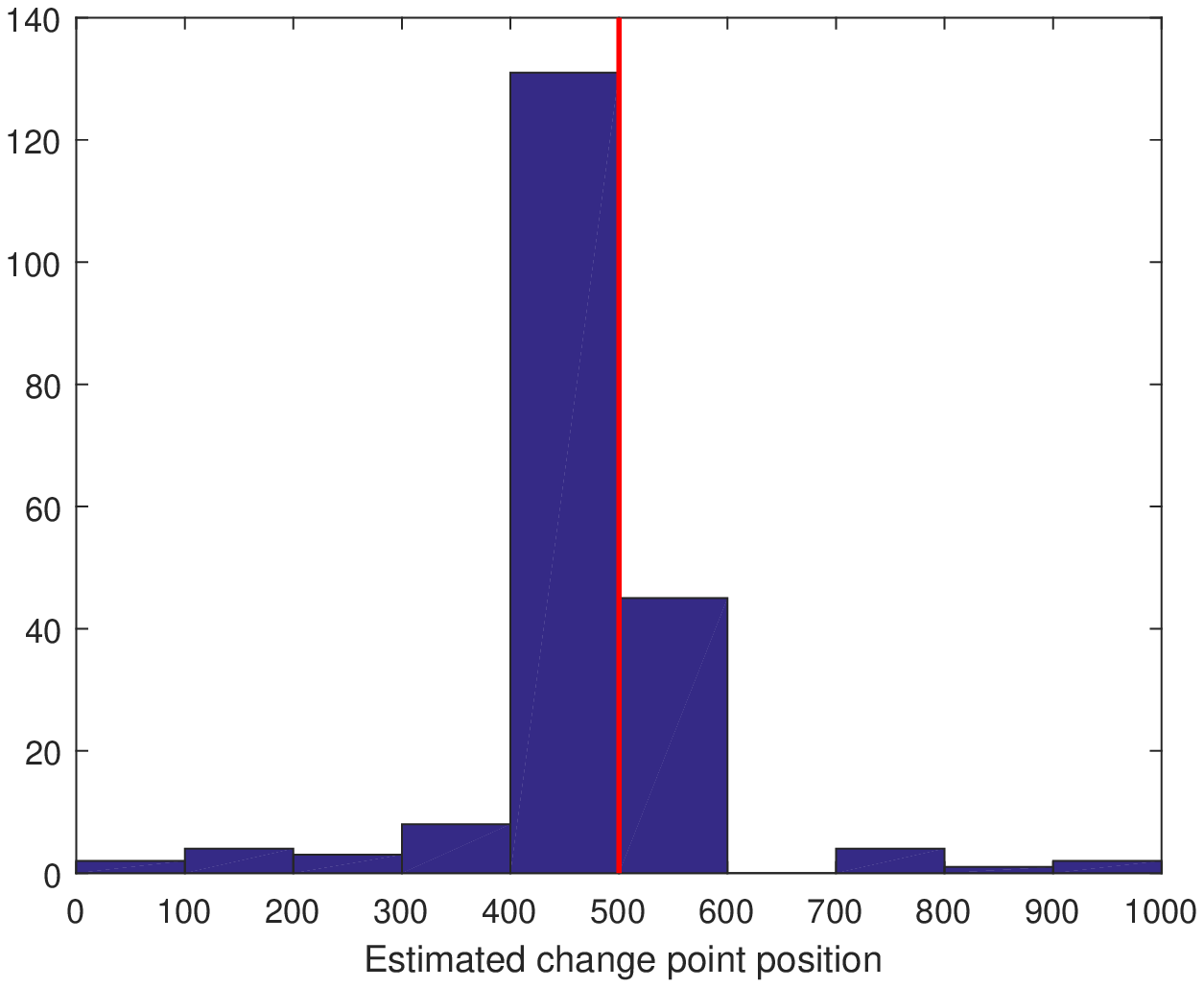}\
     \includegraphics[height = 5.5cm, width=7cm]{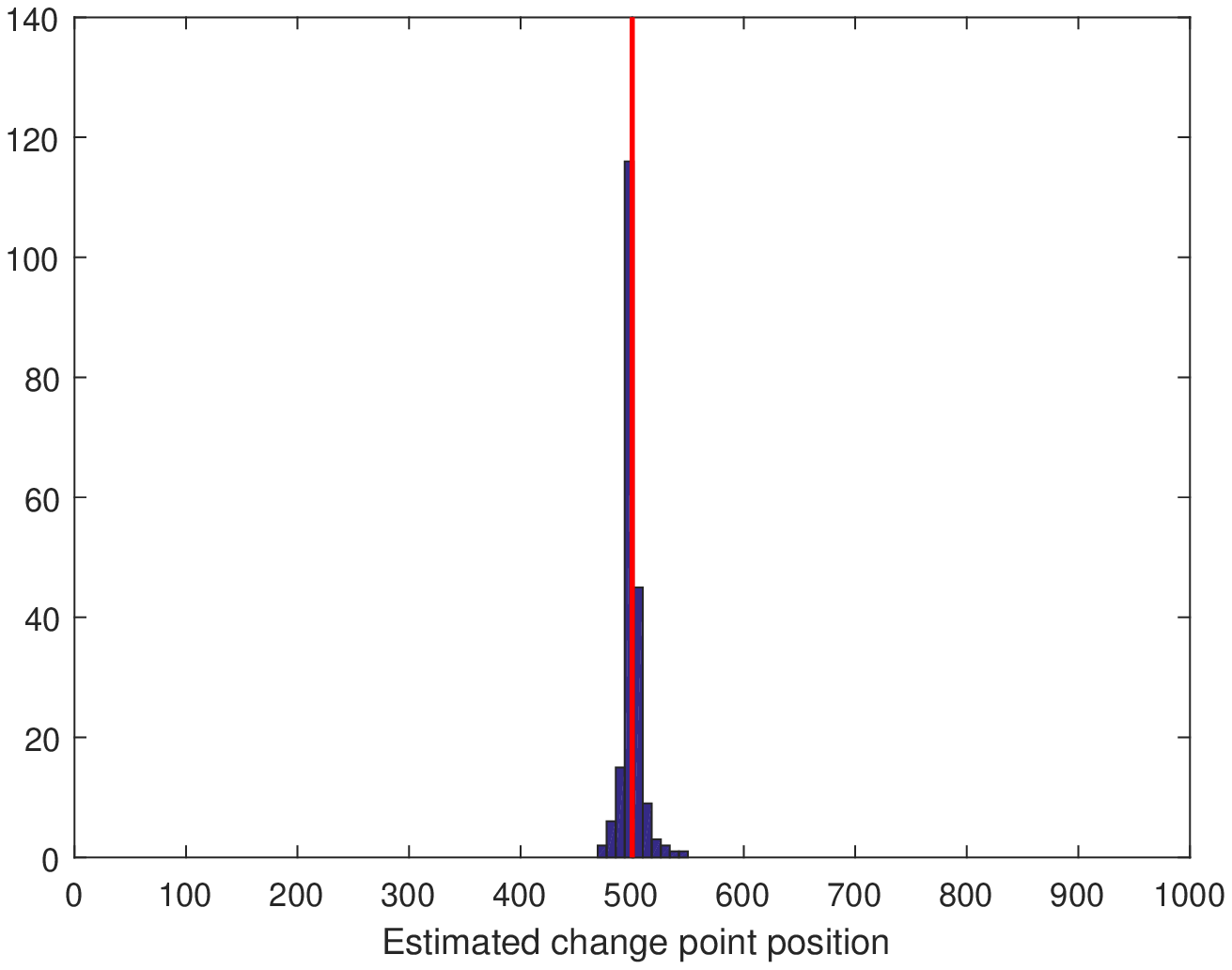}\

    \caption{\small { \it \small
    Histograms of  the  change point estimates proposed by \cite{avanbuzu2016} (left panel) and the new estimate  \eqref{yq1-8}
    proposed in this paper.  The sample size is   $n=1000$, the dimension is  $p=50$ and the ``true'' change point  is
located at $k_0=500$ (red line).}}
    \label{fig7}

\end{figure}

 In  Table \ref{Avane2} we show the simulated mean, standard deviation and mean squared error of both estimates for $200$ simulation runs.
 We observe that the estimator proposed in this paper shows a substantially better performance than the multiscale estimator introduced by
  \cite{avanbuzu2016}.  Histograms of the simulated change point positions for both methods in Figure \ref{fig7} point to the same conclusion.

\bigskip
\bigskip

{\bf Acknowledgements.}
The authors would like to thank
 M. Stein who typed parts of this manuscript with considerable technical expertise.
 We are also grateful to V.  Avanesov and N. Buzun for making the R-code
 of their procedure available to us.
The work of H. Dette   was partially supported by the Deutsche
Forschungsgemeinschaft (DFG Research Unit 1735). Guangming Pan was supported in part by by a MOE Tier 2 grant 2014-T2-2-060 and by a MOE
Tier 1 Grant RG25/14 at the Nanyang Technological University, Singapore.

\bibliographystyle{chicago}
\setlength{\bibsep}{1pt}
\bibliography{changepoint}

\newpage
\section{Proof of main results}  \label{sec5}

We first introduce some auxiliary results that are   frequently used in the proofs. Lemma \ref{lem1} is a direct conclusion of Lemma 2.7.7 in \cite{ver17}.
\begin{lem} \label{lem1}
For any $1\leq a,b\leq p$, $X_{ia}X_{ib}$ is a sub-exponential random variable, i.e. there exist positive constants $C_1, C_2 > 0$ (which do not depend on the indices $i$ and $a$) such that for every $t>0$,
\[
\pr\left(|X_{ia}X_{ib}|>t\right)\leq C_1e^{-C_2t}.
\]
\end{lem}

\begin{lem}[Corollary 2.8.3 in \cite{ver17}]\label{lem2}
Let $W_1,\cdots,W_N$ be independent, mean zero, sub-exponential random variables. Then, there exist positive constants $C_1, C_2 > 0$ such that for every $t>0$,
\[
\pr\Big\{\Big|\frac{1}{N}\sum_{i=1}^N W_i\Big| >t\Big\}\leq C_1e^{-C_2N\cdot\min (t^2,t)}.
\]
\end{lem}

\begin{lem}[Theorem 2.6.3 in \cite{ver17}]\label{lem3}
Let $W_1,\cdots,W_N$ be independent, mean zero, sub-gaussian random variables. Then, there exist positive constants $C_1, C_2$ such that for every $t>0$,
\[
\pr\Big\{\Big|\frac{1}{N}\sum_{i=1}^N W_i\Big| >t\Big\}\leq C_1e^{-C_2Nt^2}.
\]
\end{lem}

\subsection{Proof of Theorem \ref{theorem1}}\label{proof1}

Observing   the construction of the statistic $V_k$ in (\ref{yq10.1}) and (\ref{yq1-2}), we may assume without loss of generality that $\bbmu=0$.  The components of the vector $V_k$ and $D$ corresponding to the entry in the position $(a,b)$ of the matrices $\Sigma_1,\Sigma_2$, $1\leq a,b\leq p$ are given by
\begin{eqnarray}\label{yq0401.2}
V_k(a,b)&=&\frac{1}{k(k-1)}\mathop{\sum\sum}_{i\neq j\leq k}(\dot{X}_{ia}\dot{X}_{ib})(\dot{X}_{ja}\dot{X}_{jb})
+\frac{1}{(n-k)(n-k-1)}\mathop{\sum\sum}_{i\neq j>k}(\dot{X}_{ia}\dot{X}_{ib})(\dot{X}_{ja}\dot{X}_{jb})\non
&&-\frac{2}{k(n-k)}\sum_{i\leq k}\sum_{j>k}(\dot{X}_{ia}\dot{X}_{ib})(\dot{X}_{ja}\dot{X}_{jb}),\non
D(a,b)&=&\frac{1}{n-3}\sum_{k=2}^{n-2}\frac{k(n-k)}{n}V_k(a,b) = D^{(1)}(a,b)+D^{(2)}(a,b)+D^{(3)}(a,b),
\end{eqnarray}
where $\dot{X}_{ia}=X_{ia}-\bar X_a=X_{ia}-\frac{1}{n}\sum\limits_{j=1}^n X_{ja}$ by (\ref{yq10.1}) and the terms $D^{(\ell)}(a,b)$ are defined by
\begin{eqnarray*}\label{yq0401.6} \nonumber
D^{(1)}(a,b)&=&\frac{1}{n-3}\sum_{k=2}^{\lfloor \sqrt{n} \rfloor}\frac{k(n-k)}{n}V_k(a,b), \\
D^{(2)}(a,b)&=&\frac{1}{n-3}\sum_{k=\lfloor \sqrt{n} \rfloor+1}^{n-\lfloor \sqrt{n} \rfloor-1}\frac{k(n-k)}{n}V_k(a,b),\\  \nonumber
D^{(3)}(a,b)&=&\frac{1}{n-3}\sum_{k=n-\lfloor \sqrt{n} \rfloor}^{n-2}\frac{k(n-k)}{n}V_k(a,b). \nonumber
\end{eqnarray*}
 The reason
for this decomposition of  $D(a,b)$ is that all the $k$'s or $(n-k)$'s  in $D^{(2)}(a,b)$ are sufficiently large, while  both $D^{(1)}(a,b)$ and $D^{(3)}(a,b)$ only involve $(\lfloor \sqrt{n} \rfloor-1)$ terms and thus the coefficient $\frac{1}{n-3}=\frac{\lfloor \sqrt{n} \rfloor-1}{n-3}\cdot\frac{1}{\lfloor \sqrt{n} \rfloor-1} $  gives us an extra factor of order $\frac{1}{\sqrt{n}}$ in the calculations.

To be precise, let
\begin{eqnarray*}
&&\bar X_{k(a,b)}=\frac{1}{k}\sum\limits_{i=1}^k X_{ia}X_{ib}, \quad \bar Y_{k(a,b)}=\frac{1}{n-k}\sum\limits_{i=k+1}^n X_{ia}X_{ib}, \non
&&\bar X_{ka}=\frac{1}{k}\sum_{i=1}^k X_{ia},\quad\quad\quad\
\bar Y_{ka}=\frac{1}{n-k}\sum_{i=k+1}^n X_{ia},\non
&&\dot{\bar X}_{k(a,b)}=\frac{1}{k}\sum\limits_{i=1}^k \dot{X}_{ia}\dot{X}_{ib}, \quad \dot{\bar Y}_{k(a,b)}=\frac{1}{n-k}\sum\limits_{i=k+1}^n \dot{X}_{ia}\dot{X}_{ib}
\end{eqnarray*}
and
\begin{eqnarray}
\begin{split} \label{vki}
V_{k1} &=\tfrac{k(n-k)}{n}\left[\dot{\bar X}_{k(a,b)}-\dot{\bar Y}_{k(a,b)}\right]^2,\quad V_{k2}=\tfrac{k(n-k)}{n}\left[\tfrac{1}{k-1}\dot{\bar  X}_{k(a,b)}^2+\tfrac{1}{n-k-1} \dot{\bar Y}_{k(a,b)}^2\right],
\\
V_{k3} &=\tfrac{k(n-k)}{n}\Big [\tfrac{1}{k(k-1)}\sum_{i=1}^k
(\dot{X}_{ia}\dot{X}_{ib})^2+\tfrac{1}{(n-k)(n-k-1)}\sum_{i=k+1}^n
(\dot{X}_{ia}\dot{X}_{ib})^2\Big ],
\end{split}
\end{eqnarray}
then a straightforward but tedious calculation yields
\begin{equation}
\label{yq15.1}
\begin{split}
\dot{\bar X}_{k(a,b)}&=\bar X_{k(a,b)}-\bar X_{ka}\bar X_b-\bar X_{kb}\bar X_a+\bar X_a\bar X_b, \\
\dot{\bar Y}_{k(a,b)}&=\bar Y_{k(a,b)}-\bar Y_{ka}\bar X_b-\bar Y_{kb}\bar X_a+\bar X_a\bar X_b
\end{split}
\end{equation}
and
$$
\frac{k(n-k)}{n}V_k(a,b)=V_{k1}+V_{k2}-V_{k3}.
$$
With these notations we decompose the quantities $D^{(i)}(a,b)$ as follows:
\begin{equation}\label{yq0605.3}
D^{(i)}(a,b)\triangleq A^{(i)}+B^{(i)}-C^{(i)},
\end{equation}
 where
\begin{equation}
\label{yq0605.1}
\begin{split}
&A^{(1)}=\frac{1}{n-3}\sum_{k=2}^{\lfloor \sqrt{n} \rfloor}V_{k1},\quad\quad B^{(1)}=\frac{1}{n-3}\sum_{k=2}^{\lfloor \sqrt{n} \rfloor}V_{k2},\quad\quad C^{(1)}=\frac{1}{n-3}\sum_{k=2}^{\lfloor \sqrt{n} \rfloor}V_{k3},\\
&A^{(2)}=\frac{1}{n-3}\sum_{k=\lfloor \sqrt{n} \rfloor+1}^{n-\lfloor \sqrt{n} \rfloor-1}V_{k1},\quad B^{(2)}=\frac{1}{n-3}
\sum_{k=\lfloor \sqrt{n} \rfloor+1}^{n-\lfloor \sqrt{n} \rfloor-1}V_{k2},\quad C^{(2)}=\frac{1}{n-3}\sum_{k=\lfloor \sqrt{n} \rfloor+1}^{n-\lfloor \sqrt{n} \rfloor-1}V_{k3}, \\
&A^{(3)}=\frac{1}{n-3}\sum_{k=n-\lfloor \sqrt{n} \rfloor}^{n-2}V_{k1},\quad B^{(3)}=\frac{1}{n-3}\sum_{k=n-\lfloor \sqrt{n} \rfloor}^{n-2}V_{k2},\quad C^{(3)}=\frac{1}{n-3}\sum_{k=n-\lfloor \sqrt{n} \rfloor}^{n-2}V_{k3}.
\end{split}
\end{equation}

  Without loss of generality, we can assume that $\sigma_1({a,b})=\sigma_2({a,b}) =0$.
 Observing the decomposition \eqref{yq0401.2} the assertion \eqref{yq001}
follows from
\begin{equation}\label{yq0605.2}
\pr\Big\{\bigcup_{(a,b)\in \mathcal{N}}\{D^{(i)}(a,b)>c\tau\}\Big\}\rightarrow 0, \quad i=1,2,3.
\end{equation}
When we prove these results we derive exponential inequalities for all three probabilities, which directly yield the estimate in \eqref{yq001exp}.

In \eqref{yq0605.2} and hereafter in the proof, $c$ and $c_i$ $(i=1,2,\cdots)$ indicate some positive constants that may change from line to line.
According to the decomposition (\ref{yq0605.3}), it is sufficient to derive exponential inequalities, which will be used to
verify the following results:
\begin{equation}\label{yq0605.4}
p^2\cdot\pr\{A^{(i)}>c\tau\}\rightarrow 0,\quad p^2\cdot\pr\{B^{(i)}>c\tau\}\rightarrow 0,\quad p^2\cdot\pr\{C^{(i)}>c\tau\}\rightarrow 0,\quad i=1,2,3.
\end{equation}
\medskip
{\bf The case $i=1,3$.} For the index $i=1$ and $i=3$ the arguments are very similar and for the sake of brevity we only
 consider the case   $i=1$ in (\ref{yq0605.4}). For the statistic $A^{(1)}$ we find that
\begin{eqnarray}\label{yq11}
 \pr\{A^{(1)}>c\tau\}
&=& \pr\Big\{\tfrac{1}{{\sqrt{n}-1}}\sum_{k=2}^{\lfloor \sqrt{n} \rfloor}\tfrac{{\sqrt{n}-1}}{n-3}\tfrac{k(n-k)}{n}\left[\dot{\bar X}_{k(a,b)}-\dot{\bar Y}_{k(a,b)}\right]^2>c\tau\Big\}\non
&\leq& \sum_{k=2}^{\lfloor \sqrt{n} \rfloor}\pr\Big\{\tfrac{{\sqrt{n}-1}}{n-3}\tfrac{k(n-k)}{n}\left[\dot{\bar X}_{k(a,b)}-\dot{\bar Y}_{k(a,b)}\right]^2>c\tau\Big\}\non
&\leq& \sum_{k=2}^{\lfloor \sqrt{n} \rfloor}\pr\left\{\left|\dot{\bar X}_{k(a,b)}-\dot{\bar Y}_{k(a,b)}\right|>\sqrt{c\tau M_{n,k,\alpha} } \right\},
\end{eqnarray}
where we use the notation $M_{n,k,\alpha} = \frac{\sqrt{n}\cdot n}{k(n-k)}$ for the sake of a transparent notation.
Considering the components that constitute $\dot{\bar X}_{k(a,b)}$ and $\dot{\bar Y}_{k(a,b)}$ in (\ref{yq15.1}), we use  Lemma \ref{lem1} and \ref{lem2} to calculate the following probabilities to get an upper bound of (\ref{yq11})
\begin{eqnarray}\label{yq11.1}
\pr\left\{\left|\bar X_{k(a,b)}\right|>\sqrt{c\tau M_{n,k,\alpha}} \right\}&\leq& c_1e^{-c_2k\min(\tau M_{n,k,\alpha},\sqrt{\tau M_{n,k,\alpha}} )}\non
&\leq& c_1e^{-c_2\min(\tau \sqrt{n},\sqrt{\tau \sqrt{n}})},\non
\pr\left\{\left|\bar Y_{k(a,b)}\right|>\sqrt{c\tau M_{n,k,\alpha}} \right\}&\leq& c_1e^{-c_2(n-k)\min(\tau M_{n,k,\alpha},\sqrt{\tau M_{n,k,\alpha}} )}\non
&\leq&c_1 e^{-c_2\min(\tau n, \sqrt{\tau n(n-\sqrt{n})})}
\end{eqnarray}
($k=2,\ldots,  \lfloor \sqrt{n} \rfloor  $). Similarly, it follows from Lemma \ref{lem3}
\begin{eqnarray}\label{yq11.2}
&&\pr\left\{\left|\bar X_{ka}\bar X_b\right|>\sqrt{c\tau M_{n,k,\alpha}} \right\} \\
&\leq&\pr\left\{|\bar X_{ka}|> c_1\left(\tau M_{n,k,\alpha}\right)^{1/4}\right\}+\pr\left\{|\bar X_{b}|> c_2\left(\tau M_{n,k,\alpha}\right)^{1/4}\right\}\nonumber \\
&\leq&c_3\left[e^{-c_4k\sqrt{\tau M_{n,k,\alpha}}}+e^{-c_5n\sqrt{\tau M_{n,k,\alpha}}}\right]\leq c_3[e^{-c_4\sqrt{\tau \sqrt{n}}}+e^{-c_5\sqrt{\tau }n}],
\nonumber
\end{eqnarray}
for $k=2,\ldots,  \lfloor \sqrt{n} \rfloor $, and using similar arguments we obtain for
\begin{eqnarray}\label{yq11.3}
\pr\left\{\left|\bar Y_{ka}\bar X_b\right|>\sqrt{c\tau M_{n,k,\alpha}} \right\}
&\leq &
 c_3 \big [e^{-c_4\sqrt{\tau n(n-\sqrt{n})}}+e^{-c_5\sqrt{\tau }n} \big ], \\
\label{yq11.4}
 \pr\left\{\left|\bar X_{a}\bar X_b\right|>\sqrt{c\tau M_{n,k,\alpha}} \right\}
&\leq& c_3e^{-c_4n\sqrt{\tau M_{n,k,\alpha}}}\leq  c_3e^{-c_4\sqrt{\tau }n},
\end{eqnarray}
$k=2,\ldots   \lfloor \sqrt{n} \rfloor $. The same estimates (with different constants $c,c_1,c_2,\ldots$) can be derived for the remaining terms involving $\bar X_{kb}\bar X_a$ and $\bar Y_{kb}\bar X_a$.
Combining (\ref{yq15.1}), (\ref{yq11})-(\ref{yq11.4}) we obtain the upper bound
\begin{eqnarray}\label{yq16.1}
\pr\{A^{(1)}>c\tau\} &\leq& c_1\sqrt{n}\Big(e^{-c_2\min(\tau \sqrt{n},\sqrt{\tau \sqrt{n}})}+e^{-c_2\min(\tau n, \sqrt{\tau n(n-\sqrt{n})})}\non
&&+e^{-c_2\sqrt{\tau \sqrt{n}}}+e^{-c_2\sqrt{\tau n (n-\sqrt{n})}}+e^{-c_2\sqrt{\tau}n}\Big).
\end{eqnarray}
Since the smallest absolute value among the exponents in (\ref{yq16.1}) is $\sqrt{\tau \sqrt{n}}$
we have
$$
\pr\{A^{(1)}>c\tau\} \leq
 c_3 \cdot\sqrt{n}e^{-c\sqrt{\tau}n^{1/4}}
$$
 for some small positive constant $c$. Consequently, using the assumption \eqref{c2}   it follows that $p^2\cdot\pr\{A^{(1)}>c\tau\}\rightarrow 0$.

\noindent
Using similar arguments, we can investigate the other terms.
To be precise consider the statistic  $B^{(1)}$, for which we obtain the estimate
\begin{eqnarray}\label{yq12}
 \pr\{B^{(1)}>c\tau\}
&=& \pr\Big\{\tfrac{1}{n-3}\sum_{k=2}^{\lfloor \sqrt{n} \rfloor}\tfrac{k(n-k)}{n}\left[\tfrac{1}{k-1}\dot{\bar X}_{k(a,b)}^2+\tfrac{1}{n-k-1}\dot{\bar Y}_{k(a,b)}^2\right]>c\tau\Big\}\\
&\leq& \sum_{k=2}^{\lfloor \sqrt{n} \rfloor}\pr\left\{\tfrac{{\sqrt{n}-1}}{n-3}\tfrac{k(n-k)}{n}\left[\tfrac{1}{k-1}\dot{\bar X}_{k(a,b)}^2+\tfrac{1}{n-k-1}\dot{\bar Y}_{k(a,b)}^2\right]>c\tau\right\}\non
&\leq& \sum_{k=2}^{\lfloor \sqrt{n} \rfloor} \Big [
\pr\Big \{\big |\dot{\bar X}_{k(a,b)}\big |>\sqrt{c\tau M^{(1)}_{n,k,\alpha} }\Big\}+
\pr\Big \{\big |\dot{\bar Y}_{k(a,b)}\big |>\sqrt{c\tau M^{(2)}_{n,k,\alpha} }\Big\} \Big ] , \nonumber
\end{eqnarray}
where the quantities $M^{(1)}_{n,k,\alpha}$ and $M^{(2)}_{n,k,\alpha}$ are defined by
$$
M^{(1)}_{n,k,\alpha} = \frac{\sqrt{n}\cdot n}{(n-k)}~ \mbox{ and } ~  M^{(2)}_{n,k,\alpha}=  \frac{\sqrt{n}\cdot n}{k},
$$
respectively. Again we  consider  the components that constitute $\dot{\bar X}_{k(a,b)}$ and $\dot{\bar Y}_{k(a,b)}$ in (\ref{yq15.1})
separately and obtain by an application of   Lemma \ref{lem1} and \ref{lem2} the following estimates
(for $ k=2 , \ldots ,   \lfloor \sqrt{n} \rfloor  $)
\begin{eqnarray*}
\pr\Big \{\left|\bar X_{k(a,b)}\right|>\sqrt{c\tau M^{(1)}_{n,k,\alpha}} \Big \}&\leq&
c_1e^{-c_2k\min(\tau M^{(1)}_{n,k,\alpha},\sqrt{\tau M^{(1)}_{n,k,\alpha}} )} \leq  c_1 e^{-c_2\min(\tau \sqrt{n},\sqrt{\tau \sqrt{n}})},\non
\pr\Big \{\left|\bar Y_{k(a,b)}\right|>\sqrt{c\tau M^{(2)}_{n,k,\alpha}} \Big \}&\leq&
c_1e^{-c_2(n-k)\min(\tau M^{(2)}_{n,k,\alpha},\sqrt{\tau M^{(2)}_{n,k,\alpha}} )}\non
&\leq& c_1e^{-c_2\min(\tau n(n-\sqrt{n}), \sqrt{\tau n(n-\sqrt{n})^2})}.
\end{eqnarray*}
Similarly,  Lemma \ref{lem3} gives for   $ k=2 , \ldots , \lfloor \sqrt{n} \rfloor$
\begin{eqnarray*}
\pr\Big\{\left|\bar X_{ka}\bar X_b\right|>\sqrt{c\tau M^{(1)}_{n,k,\alpha}} \Big\}
&\leq&  c_3\Big[e^{-c_4k\sqrt{\tau M^{(1)}_{n,k,\alpha}}}+e^{-c_5n\sqrt{\tau M^{(1)}_{n,k,\alpha}}}\Big]\\
&\leq& c_3   \left[e^{-c_4\sqrt{\tau \sqrt{n}}}+e^{-c_5n\sqrt{\tau \sqrt{n}}}\right],\non
\pr\Big\{\left|\bar Y_{ka}\bar X_b\right|>\sqrt{c\tau M^{(2)}_{n,k,\alpha}} \Big\}
&\leq&
c_3\Big[e^{-c_4(n-k)\sqrt{\tau M^{(2)}_{n,k,\alpha}}}+e^{-c_5n\sqrt{\tau M^{(2)}_{n,k,\alpha}}}\Big] \\
&\leq&
c_3\left[e^{-c_4(n-\sqrt{n})\sqrt{\tau n}}+e^{-c_5n\sqrt{\tau n}}\right],\non
\pr\Big\{\left|\bar X_{a}\bar X_b\right|>\sqrt{c\tau M^{(1)}_{n,k,\alpha}} \Big\}
&\leq & c_3e^{-c_4n\sqrt{\tau M^{(1)}_{n,k,\alpha}}}\leq c_3e^{-c_4n\sqrt{\tau \sqrt{n}}},\non
\pr\Big\{\left|\bar X_{a}\bar X_b\right|>\sqrt{c\tau M^{(2)}_{n,k,\alpha}} \Big\}
&\leq & c_3e^{-c_4n\sqrt{\tau M^{(2)}_{n,k,\alpha}}}\leq c_3e^{-c_4n\sqrt{\tau n}}.
\end{eqnarray*}
Thus, observing that the   smallest absolute value among the exponents in these estimates  is  given by $\sqrt{\tau \sqrt{n}}$,
 an upper bound for the probability in (\ref{yq12}) is obtained as
$$
 \pr\{B^{(1)}>c\tau\}\leq c_1  \sqrt{n} e^{-c\sqrt{\tau} n^{1/4}}
$$
 for some small positive constant $c$. Consequently, observing Assumption \ref{cond5} we have $p^2\pr\{B^{(1)}>c\tau\}\rightarrow 0$  as $n,p \to \infty$.

Finally the term  $C^{(1)}$ is estimated as follows
\begin{eqnarray*}\label{yq13}
\pr\{C^{(1)}>c\tau\}
&\leq&\sum_{k=2}^{\lfloor \sqrt{n} \rfloor}\pr\Big \{\tfrac{{\sqrt{n}-1}}{n-3}\tfrac{k(n-k)}{n}\Big [\tfrac{1}{k(k-1)}\sum_{i=1}^k
(\dot{X}_{ia}\dot{X}_{ib})^2+\tfrac{1}{(n-k)(n-k-1)}\sum_{i=k+1}^n
(\dot{X}_{ia}\dot{X}_{ib})^2\Big ]>c\tau\Big \}\non
&\leq&\sum_{k=2}^{\lfloor \sqrt{n} \rfloor} \Big[
\pr\Big  \{\tfrac{1}{k}\sum_{i=1}^k(\dot{X}_{ia}\dot{X}_{ib})^2>c_1\tau\tfrac{n\sqrt{n}}{n-k} \Big\}+
 \pr\Big \{\tfrac{1}{n-k}\sum_{i=k+1}^n(\dot{X}_{ia}\dot{X}_{ib})^2>c_2\tau \tfrac{ n\sqrt{n}}{k} \Big \}
 \Big] \non
&\leq&\sum_{k=2}^{\lfloor \sqrt{n} \rfloor} \Big[
k\pr\Big \{|\dot{X}_{ia}\dot{X}_{ib}|>\sqrt{c_1\tau \sqrt{n}}\Big\}+ (n-k)
\pr\Big \{|\dot{X}_{ia}\dot{X}_{ib}|>\sqrt{c_2\tau n} \Big\} \Big] .
\end{eqnarray*}
Observing that
$\dot{X}_{ia}\dot{X}_{ib}=X_{ia}X_{ib}-X_{ia}\bar X_{b}-X_{ib}\bar X_{a}+\bar X_{a}\bar X_b
$,
we obtain from  Lemma \ref{lem1} and  \ref{lem3} the estimates
\begin{eqnarray*}
  k\pr\Big\{|X_{ia}X_{ib}|>\sqrt{c_1\tau \sqrt{n}} \Big\} &\leq&
c \sqrt{n}e^{-c_4\sqrt{\tau \sqrt{n}}},\non
 (n-k)\pr\Big\{|X_{ia}X_{ib}|>\sqrt{c_2\tau n} \Big\} &\leq&
cn e^{-c_6\sqrt{\tau n}}, \non
k\pr\Big\{|X_{ia}\bar X_b|>\sqrt{c_1\tau \sqrt{n}} \Big\} &\leq&
 c\sqrt{n}\left[e^{-c_7\sqrt{\tau \sqrt{n}}}+e^{-c_8n\sqrt{\tau \sqrt{n}}}\right],\non
 (n-k)\pr\left\{|X_{ia}\bar X_b|>\sqrt{c_2\tau n} \right\}
 &\leq&  cn\left[e^{-c_7\sqrt{\tau n}}+e^{-c_8n\sqrt{\tau n}}\right], \non
  k\pr\Big\{|\bar X_a\bar X_{b}|>\sqrt{c_1\tau \sqrt{n}} \Big\}
 &\leq& c \sqrt{n}e^{-c_7n\sqrt{\tau \sqrt{n}}},\non
 (n-k)\pr\left\{|\bar X_a\bar X_{b}|>\sqrt{c_2\tau n} \right\}
 &\leq& c n e^{-c_7n\sqrt{\tau n}},
\end{eqnarray*}
whenever $k=2, \ldots  , \lfloor \sqrt{n} \rfloor$.
Summarizing we have
\begin{eqnarray*}\label{yq20.1}
 \pr\{C^{(1)}>c\tau\}&\leq & c_1 \Big[n(e^{-c_2\sqrt{\tau \sqrt{n}}}+e^{-c_2n\sqrt{\tau \sqrt{n}}})+
n\sqrt{n}(e^{-c_2\sqrt{\tau n}}+e^{-c_2n\sqrt{\tau n}})\Big]\non
&\leq &c_3ne^{-c\sqrt{\tau \sqrt{n}}}
\end{eqnarray*}
for some small positive constant $c$. Now assumption \eqref{c2}  implies   $p^2\pr\{C^{(1)}>c\tau\}\rightarrow 0$, if
\begin{equation}\label{yq0606.1}
p^2ne^{-c \sqrt{\tau}n^{1/4}} \rightarrow 0,
\end{equation}
 and therefore   the proof
of (\ref{yq0605.4})  in the case $i=1$ is completed (the case $i=3$ follows by exactly the same arguments).

\noindent
{\bf The case $i=2$.}
For the term $A^{(2)}$, we get that
\begin{eqnarray*}\label{yq21}
\pr\{A^{(2)}>c\tau\}
&\leq&\sum_{k=\lfloor \sqrt{n} \rfloor+1}^{n-\lfloor \sqrt{n} \rfloor-1} \Big [
\pr\Big \{\big|\dot{\bar X}_{k(a,b)}\big|>\sqrt{c_1\tau M_{n,k}}\Big \}+ \pr\Big \{\big |\dot{\bar Y}_{k(a,b)}\big |>\sqrt{c_2\tau M_{n,k} }\Big\} \Big] , ~~~~
\end{eqnarray*}
where we use the notation  $M_{n,k}= \frac{n}{k(n-k)}$. Similar calculations as given  for the term  $A^{(1)}$ give for the  summands  in
the representation (\ref{yq15.1})  of  $\dot{\bar X}_{k(a,b)}$ and $\dot{\bar Y}_{k(a,b)}$ the estimates (here we use the fact that for $i=2$ we have $k=\lfloor \sqrt{n} \rfloor +1, \ldots, n-\lfloor \sqrt{n} \rfloor-1$)
\begin{eqnarray*}
 \pr\left\{\left|\bar X_{k(a,b)}\right|>\sqrt{c\tau M_{n,k}} \right\}
&\leq&   c_1e^{-c_2k\min(\tau M_{n,k},\sqrt{\tau M_{n,k}} )} \leq
c_1  e^{-c_2\min(\tau ,\sqrt{\tau \sqrt{n}})},\non
 \pr\left\{\left|\bar Y_{k(a,b)}\right|>\sqrt{c\tau M_{n,k}} \right\}
&\leq&  c_1e^{-c_2(n-k)\min(\tau M_{n,k},\sqrt{\tau M_{n,k}} )} \leq
c_1  e^{-c_2\min(\tau , \sqrt{\tau \sqrt{n}})} ,\non
 \pr\left\{\left|\bar X_{ka}\bar X_b\right|>\sqrt{c\tau M_{n,k}} \right\}
&\leq& c_3\left[e^{-c_4k\sqrt{\tau M_{n,k}}}+e^{-c_5n\sqrt{\tau M_{n,k}}}\right]\leq c_3\left[e^{-c_4\sqrt{\tau \sqrt{n}}}+e^{-c_5\sqrt{\tau n}}\right], \non
 \pr\left\{\left|\bar Y_{ka}\bar X_b\right|>\sqrt{c\tau M_{n,k}} \right\}
&\leq& c_3\left[e^{-c_4(n-k)\sqrt{\tau M_{n,k}}}+e^{-c_5n\sqrt{\tau M_{n,k}}}\right]\leq c_3\left[e^{-c_4\sqrt{\tau \sqrt{n}}}+e^{-c_5\sqrt{\tau n}}\right],\non
 \pr\left\{\left|\bar X_a\bar X_b\right|>\sqrt{c\tau M_{n,k}} \right\}
&\leq & c_3e^{-c_4n\sqrt{\tau M_{n,k}}}\leq  c_3 e^{-c_4\sqrt{\tau n}},
\end{eqnarray*}
and consequently we obtain
$$
p^2\cdot\pr\{A^{(2)}>c\tau\} \leq  c_1  n p^2 \max \big \{ e^{-c\tau } ,  e^{-c\sqrt{\tau \sqrt{n}}} \big\},
$$
 for some small positive constant $c$, which converges to $0$ under the stated assumptions \eqref{c1} and \eqref{c2}.

For the term $B^{(2)}$ we have
\begin{eqnarray}\label{yq22}
\pr\{B^{(2)}>c\tau\}
&\leq&\sum_{k=\lfloor \sqrt{n} \rfloor+1}^{n-\lfloor \sqrt{n} \rfloor-1}
\Big [ \pr\Big \{\big  |\dot{\bar X}_{k(a,b)}\big |>\sqrt{c_1\tau\tfrac{n}{n-k}}\Big \}+
\pr\Big  \{\big |\dot{\bar Y}_{k(a,b)}\big |>\sqrt{c_2\tau\tfrac{n}{k}}\Big  \} \Big] , ~~~
\end{eqnarray}
where the two probabilities can be bounded    taking into account the representation  (\ref{yq15.1}) and the
estimates
\begin{eqnarray*}
&&
\pr\Big \{\big  |{ \bar X}_{k(a,b)} \big |>\sqrt{\tfrac{c_1\tau n}{n-k}}\Big \}
\leq
c_1e^{-c_2k\min(\frac{\tau n}{n-k},\sqrt{\frac{\tau n}{n-k}})}
\leq c_1 e^{-c_2\min(\tau \sqrt{n},\sqrt{\tau n})},\non
&&\pr\left\{\left|{\bar Y}_{k(a,b)}\right|>\sqrt{\tfrac{c\tau n}{k}}\right\}\leq
c_1e^{-c_2(n-k)\min(\frac{\tau n}{k},\sqrt{\frac{\tau n}{k}})}
\leq c_1e^{-c_2\min(\tau \sqrt{n},\sqrt{\tau n})}, \non
&&\pr\left\{\left|\bar X_{ka}\bar X_b\right|>\sqrt{\tfrac{c\tau n}{n-k}}\right\}\leq
c_3\left[e^{-c_4k\sqrt{\frac{\tau n}{n-k}}}+e^{-c_5n\sqrt{\frac{\tau n}{n-k}}}\right]\leq c_3
[e^{-c_4\sqrt{\tau n}}+e^{-c_5n\sqrt{\tau}}],\non
&&
\pr\left\{\left|\bar Y_{ka}\bar X_b\right|>\sqrt{\tfrac{c\tau n}{k}}\right\}\leq c_3\left[e^{-c_4(n-k)\sqrt{\frac{\tau n}{k}}}+e^{-c_5n\sqrt{\frac{\tau n}{k}}}\right]\leq c_3 [e^{-c_4\sqrt{\tau n}}+e^{-c_5n\sqrt{\tau}}],\non
&&  \pr\left\{\left|\bar X_{a}\bar X_b\right|>\sqrt{\tfrac{c\tau n}{n-k}}\right\}
\leq  c_3e^{-c_4n\sqrt{\frac{\tau n}{n-k}}}\leq c_3  e^{-c_4n\sqrt{\tau}},\non
&& \pr\left\{\left|\bar X_{a}\bar X_b\right|>\sqrt{\tfrac{c\tau n}{k}}\right\}\leq c_3 e^{-c_4n\sqrt{\tau}}
\end{eqnarray*}
for $k=\lfloor \sqrt{n} \rfloor+1 , \lfloor \sqrt{n} \rfloor+2 , \ldots , n-\lfloor \sqrt{n} \rfloor-1 $.
Therefore, we obtain as an  upper bound for the probability in  (\ref{yq22})
\begin{eqnarray*}\label{yq20.5}
p^2\cdot\pr\{B^{(2)}>c\tau\}&\leq& c_1p^2n\Big(e^{-c_2\min(\tau \sqrt{n},\sqrt{\tau n})}+e^{-c_2\sqrt{\tau n}}+e^{-c_2n\sqrt{\tau}}\Big)\non
&=&c_1p^2n\Big(e^{-c_2\sqrt{n}\sqrt{\tau}}+e^{-c_2\sqrt{\tau n}}+e^{-c_2n\sqrt{\tau}}\Big)= o(1),
\end{eqnarray*}
 by assumption \eqref{c2}.

\noindent
For the term $C^{(2)}$, we observe (note that $n/(n-k)>1$ and $n/k>1$)
\begin{eqnarray} \label{yq23}
\pr\{C^{(2)}>c\tau\}
&\leq&\sum_{k=\lfloor \sqrt{n} \rfloor+1}^{n-\lfloor \sqrt{n} \rfloor-1}\Big [ \pr
\Big
\{\tfrac{1}{k}\sum_{i=1}^k(\dot{X}_{ia}\dot{X}_{ib})^2>c_1\tau \tfrac{n}{n-k} \Big  \}+ \pr\Big \{\tfrac{1}{n-k}\sum_{i=k+1}^n(\dot{X}_{ia}\dot{X}_{ib})^2>c_2\tau \tfrac{n}{k} \Big \}  \Big ]\non
&\leq& \sum_{k=\lfloor \sqrt{n} \rfloor+1}^{n-\lfloor \sqrt{n} \rfloor-1}k\pr\left\{|\dot{X}_{ia}\dot{X}_{ib}|>\sqrt{c_1\tau} \right\}
\nonumber
\\
& + &
\sum_{k=\lfloor \sqrt{n} \rfloor+1}^{n-\lfloor \sqrt{n} \rfloor-1}(n-k)\pr\left\{|\dot{X}_{ia}\dot{X}_{ib}|>\sqrt{c_2\tau} \right\},
\end{eqnarray}
where the two probabilities can be bounded using the representation
$\dot{X}_{ia}\dot{X}_{ib}=X_{ia}X_{ib}-X_{ia}\bar X_{b}-X_{ib}\bar X_{a}+\bar X_{a}\bar X_b$. This gives for $k= \lfloor \sqrt{n} \rfloor+1,  \ldots , n-\lfloor \sqrt{n} \rfloor-1$ the estimates
\begin{eqnarray*}
 \pr\left\{|{X}_{ia}{X}_{ib}|>c\sqrt{\tau} \right\}&\leq& c_1 e^{-c_2\sqrt{\tau}},\non
 \pr\left\{|{X}_{ia}\bar{X}_{b}|>c\sqrt{\tau} \right\}&\leq&  \left[\pr\left\{|{X}_{ia}|>c_1\tau^{1/4} \right\}+\pr\left\{|\bar{X}_{b}|>c_2\tau^{1/4} \right\}\right]
\leq c_3 \Big[e^{-c_4\sqrt{\tau}}+e^{-c_5n\sqrt{\tau}}\Big],\non
 \pr\left\{|\bar{X}_{a}\bar{X}_{b}|>c\sqrt{\tau} \right\}&\leq&   \left[\pr\left\{|\bar{X}_{a}|>c_1\tau^{1/4} \right\}+\pr\left\{|\bar{X}_{b}|>c_2\tau^{1/4} \right\}\right]\leq c_3 e^{-c_4n\sqrt{\tau}}.
\end{eqnarray*}
Observing that $\sum^{n-\lfloor \sqrt{n} \rfloor-1}_{k=\lfloor \sqrt{n} \rfloor+1} k \leq n^2$, $\sum_{k=\lfloor \sqrt{n} \rfloor+1}^{n-\lfloor \sqrt{n} \rfloor-1}(n-k)\leq n^2$  we can bound  the probability on the left hand side of (\ref{yq22}) by
$$
 \pr\{C^{(2)}>c\tau\}\leq c_1 n^2\Big(e^{-c_2\sqrt{\tau}}+e^{-c_3n\sqrt{\tau}}\Big),
$$
which vanishes asymptotically by assumption \eqref{c3}.

Therefore, we have established \eqref{yq0605.4} for all indices $i = 1,2,3$ which implies \eqref{yq0605.2}, and the assertion of Theorem \ref{theorem1} follows. A careful inspection of our arguments shows that we have also established the estimate \eqref{yq001exp} in Theorem \ref{theorem1}.

\subsection{Proof of Theorem \ref{theorem1a}}

Recall that  the set $\mathcal{P}$ is the set of indices corresponding to elements with
$|\sigma_1({a,b})-\sigma_2({a,b}) |>\lambda$, where   $\lambda$ satisfies (\ref{yq0531.1}), and  note that the assertion \eqref{yq002} is equivalent to
$$
\pr\Big\{\bigcup_{(a,b)\in \mathcal{P}}\{D(a,b)\leq\tau\}\Big\}\rightarrow 0.
$$
For a proof of this statement we derive several  exponential inequalities which directly yield the estimate in \eqref{yq002exp}.
For this purpose we  introduce the decomposition
\[
D(a,b)=\frac{1}{n-3}\sum_{k=2}^{n-2}[V_{k1}+V_{k2}-V_{k3}] =  A+B-C,\]
 where the quantities $A$, $B$ and $C$ are given by
\[
A=\frac{1}{n-3}\sum_{k=2}^{n-2}V_{k1},\quad B=\frac{1}{n-3}\sum_{k=2}^{n-2}V_{k2},\quad C=\frac{1}{n-3}\sum_{k=2}^{n-2}V_{k3},
\]
respectively, and the statistics $V_{k1}, V_{k2}$ and $V_{k_3}$ have been defined in \eqref{vki}.
Observing the inclusion
 \begin{eqnarray*} \label{inclusion}
 \{D(a,b)\leq\tau\}\  \subset \{A\leq3\tau\}\cup\{|B-C|\geq 2\tau\}  \subset  \{A\leq3\tau\}\cup\{B\geq \tau\}\cup\{C\geq \tau\}
\end{eqnarray*}
  the assertion of Theorem \ref{theorem1a} follows  by deriving exponential inequalities for the probabilities for these three events, which are used to prove
\begin{equation}\label{yq0606.7}
p^2\cdot\pr\{A\leq3\tau\}\rightarrow 0,\quad p^2\cdot\pr\{B\geq \tau\}\rightarrow 0,\quad p^2\cdot\pr\{C\geq \tau\}\rightarrow 0.
\end{equation}
We now investigate the three probabilities in (\ref{yq0606.7}) separately. First, for the last two terms $B$ and $C$,
from (\ref{yq0605.1}), it is easy to see that $B=B^{(1)}+B^{(2)}+B^{(3)}$ and $C=C^{(1)}+C^{(2)}+C^{(3)}$, where - as in the proof of Theorem \ref{theorem1} - the indices $(1)$, $(2)$ and $(3)$ correspond to the summation with respect to the sets $\{ 2,\ldots,\lfloor \sqrt{n} \rfloor \}, \{\lfloor \sqrt{n} \rfloor +1, \ldots, n-\lfloor \sqrt{n} \rfloor -1\}$ and $\{n-\lfloor \sqrt{n} \rfloor,\ldots,n-2  \}$, respectively.
Moreover,  using similar arguments as given in the proof of Theorem \ref{theorem1}, the following results can be established.
\begin{eqnarray}\label{h1}
p^2\cdot \pr\{B^{(i)}>c\tau\} &\leq &c_1  p^2\sqrt{n} e^{-c\sqrt{\tau} n^{1/4}}, i=1,3,\quad p^2\cdot\pr\{B^{(2)}>c\tau\}\leq c_1p^2ne^{-c_2\sqrt{\tau n}},\non
 p^2\cdot\pr\{C^{(i)}>c\tau\} & \leq &c_3p^2ne^{-c\sqrt{\tau}n^{1/4}}, i=1,3,\quad
p^2\cdot\pr\{C^{(2)}>c\tau\}\leq c_1p^2 n^2e^{-c_2\sqrt{\tau}}. ~~~~
\end{eqnarray}
The only difference when analyzing the above probabilities lies in the expectation of $X_{ia}X_{ib}$, $\E(X_{ia}X_{ib})$, which is not necessarily zero now. But this does not affect the proof  of \eqref{h1} since $\E(X_{ia}X_{ib})$  is always bounded by Assumption \ref{cond1}. Hence
\[
p^2\cdot\pr\{B\geq \tau\}\rightarrow 0,\quad p^2\cdot\pr\{C\geq \tau\}\rightarrow 0.
\]
In order to show the remaining exponential equation and prove the  assertion  $p^2\cdot\pr\{A\leq3\tau\}\rightarrow 0$, we note that $\sigma_1({a,b})\not = \sigma_2({a,b}) $, and a straightforward calculation gives
\begin{eqnarray*}
 \E(\bar X_{k(a,b)}-\bar Y_{k(a,b)})=\left\{
  \begin{array}{l l}
    \frac{n-k_0}{n-k}(\sigma_1({a,b})-\sigma_2({a,b}))  & : k\leq k_0,\\
    \frac{k_0}{k}(\sigma_1({a,b})-\sigma_2({a,b}) ) & : k>k_0.
  \end{array}
\right.
\end{eqnarray*}
Let
\begin{eqnarray*}
A_k=\dot{\bar X}_{k(a,b)}-\dot{\bar Y}_{k(a,b)}-\E({\bar X}_{k(a,b)}-{\bar Y}_{k(a,b)}),\quad B_k=\E({\bar X}_{k(a,b)}-{\bar Y}_{k(a,b)}),
\end{eqnarray*}
and define $b_{n,k}\triangleq \frac{k(n-k)}{n}$. Then one can observe that
\begin{equation}\label{yq11.5}
|B_k|\geq \min\left\{\tfrac{n-k_0}{n},\tfrac{k_0}{n} \right\}\cdot|\sigma_1({a,b})-\sigma_2({a,b})|
\end{equation}
and
\begin{eqnarray*}
\tfrac{1}{n-3}\sum_{k=2}^{n-2}b_{n,k}B_k^2 &\geq & \min\left\{\tfrac{(n-k_0)^2}{n^2},\tfrac{k_0^2}{n^2} \right\}\cdot  n(\sigma_1({a,b})-\sigma_2({a,b}) )^2\cdot\tfrac{1}{n-3}\sum_{k=2}^{n-2}\tfrac{k(n-k)}{n^2}\non
&\geq&\min\left\{\tfrac{(n-k_0)^2}{n^2},\tfrac{k_0^2}{n^2} \right\}\cdot\tfrac{ n(\sigma_1({a,b})-\sigma_2({a,b}) )^2}{6}
\geq\min\left\{\tfrac{(n-k_0)^2}{n^2},\tfrac{k_0^2}{n^2} \right\}\cdot\tfrac{ n\lambda^2}{6}.
\end{eqnarray*}
Consequently, by assumption \eqref{yq0531.1} (with a sufficiently large constant $C$) we  obtain the estimate
$$
3\tau\leq \frac{1}{2(n-3)}\sum\limits_{k=2}^{n-2}b_{n,k}B_k^2.
$$
On the other hand, we have by the definition of $A$, $A_k$ and $B_k$
\begin{eqnarray}\label{yq0402.6}
\pr\{A\leq3\tau\}
&=& \pr\Big\{\tfrac{1}{n-3}\sum_{k=2}^{n-2}b_{n,k}\left[A_k+B_k\right]^2\leq 3\tau \Big\}\non
&\leq& \pr\Big\{\tfrac{2}{n-3}\sum_{k=2}^{n-2}b_{n,k}A_kB_k\leq 3\tau-\tfrac{1}{n-3}\sum_{k=2}^{n-2}b_{n,k}B_k^2 \Big\}\non
&\leq& \pr\Big\{\tfrac{2}{n-3}\sum_{k=2}^{n-2}b_{n,k}A_kB_k\leq -\tfrac{1}{2(n-3)}\sum_{k=2}^{n-2}b_{n,k}B_k^2 \Big\}\non
&\leq& \pr\Big\{\tfrac{2}{n-3}\Big|\sum_{k=2}^{n-2}b_{n,k}A_kB_k\Big|\geq \tfrac{1}{2(n-3)}\Big|\sum_{k=2}^{n-2}b_{n,k}B_k^2\Big| \Big\}\non
&\leq& n\cdot\pr\left\{\left|b_{n,k}A_k\right|\geq \tfrac{c}{|B_k|}\min\left\{\tfrac{(n-k_0)^2}{n^2},\tfrac{k_0^2}{n^2} \right\}\cdot n(\sigma_1({a,b})-\sigma_2({a,b}))^2 \right\}\non
&=& n\cdot\pr\left\{\left|A_k\right|\geq c\phi\right\},
\end{eqnarray}
where $\phi=\frac{1}{|B_k|}\frac{1}{b_{n,k}}\min\left\{\frac{(n-k_0)^2}{n^2},\frac{k_0^2}{n^2} \right\}\cdot n(\sigma_1({a,b})-\sigma_2({a,b}) )^2 $. In order to investigate the probability $\pr\left\{\left|A_k\right|\geq c\phi\right\}$ in (\ref{yq0402.6})   note that $\frac{1}{|B_k|}\geq\frac{1}{|\sigma_1({a,b})-\sigma_2({a,b}) |}$ by (\ref{yq11.5}). If
\[|\sigma_1({a,b})-\sigma_2({a,b}) |\geq \lambda\geq C\sqrt{\tfrac{\tau}{n}}\max\left\{\tfrac{n^2}{(n-k_0)^2},\tfrac{n^2}{k_0^2}\right\},
 \]
 it is therefore easy to see that
\[
\phi\geq \tfrac{n}{k(n-k)}\min\left\{\tfrac{(n-k_0)^2}{n^2},\tfrac{k_0^2}{n^2} \right\}\cdot n|\sigma_1({a,b})-\sigma_2({a,b}) |\geq
C \tfrac{n}{k(n-k)}\sqrt{n\tau}.
\]
Observing the decomposition (\ref{yq15.1}), the term $A_k$ can be written as
\begin{eqnarray*}
A_k
&=&[\bar X_{k(a,b)}-\bar Y_{k(a,b)}-\E(\bar X_{k(a,b)}-\bar Y_{k(a,b)})]-\bar X_{ka}\bar X_b-\bar X_{kb}\bar X_a+\bar Y_{ka}\bar X_b+\bar Y_{kb}\bar X_a,
\end{eqnarray*}
and by Lemma \ref{lem2} and Lemma \ref{lem3} we obtain
\begin{eqnarray}\label{yq0608.2}
&&\pr\left\{\left|A_k\right|\geq c\phi\right\}\non
&\leq&\pr\left\{\left|\bar X_{k(a,b)}-\E\bar X_{k(a,b)}\right|\geq c_1\phi\right\}
+\pr\left\{\left|\bar Y_{k(a,b)}-\E\bar Y_{k(a,b)}\right|\geq c_2\phi\right\}\non
&&+2\pr\left\{\left|\bar X_{ka}\right|\geq c_3\sqrt{\phi}\right\}+2\pr\left\{\left|\bar X_{b}\right|\geq c_4\sqrt{\phi}\right\}+2\pr\left\{\left|\bar Y_{ka}\right|\geq c_3\sqrt{\phi}\right\}+2\pr\left\{\left|\bar X_{b}\right|\geq c_4\sqrt{\phi}\right\}\non
&\leq&c_5\Big[e^{-c_6k\min\{\phi^2,\phi\}}+e^{-c_6(n-k)\min\{\phi^2,\phi\}}+e^{-c_6k\phi}+e^{-c_6(n-k)\phi}+e^{-c_6n\phi}\Big]\non
&\leq&c_5\Big[e^{-c_6\min\{\tau,\sqrt{n\tau}\}}+e^{-c_6\min\{\tau,\sqrt{n\tau}\}}+e^{-c_6\sqrt{n\tau}}+e^{-c_6\sqrt{n\tau}}+e^{-c_6\sqrt{n\tau}}\Big] = o(1),
\end{eqnarray}
where the last estimate follows from \eqref{c1} and \eqref{c3}.
Combining (\ref{yq0402.6}) with (\ref{yq0608.2}), it follows that $p^2\cdot\pr\{A\leq3\tau\}\rightarrow 0$, which completes the proof of Theorem \ref{theorem1a}.

\subsection{Proof of Theorem \ref{theorm2}}\label{proof}
Recall the definition of the statistic $U_n(k)$ in \eqref{yq1-7}
  and the definition of the change point estimator $\hat k$ in \eqref{yq1-8}.
Let $\widetilde{\bbx}_i$ indicate the  $m$-dimensional subvector of $\vv(\bbx_i\bbx_i^T)$ corresponding to the components   in the set ${\cal D}_\tau$ in \eqref{yq1-4}. Obviously,
\begin{equation}\label{yq0531.3}
\pr\Big\{\Big|\tfrac{\hat k}{k_0}-1\Big|\geq\epsilon\Big\}\leq
\pr\Big\{\hat k\geq(1+\epsilon)k_0\Big\}+\pr\Big\{\hat k\leq (1-\epsilon)k_0\Big\},
\end{equation}
and we will derive exponential bounds for the  two terms on the right-hand side
to prove that the probability vanishes asymptotically. We only consider the first term because the
second term can be handled similarly. It is sufficient to show that
\begin{eqnarray*}
P\Big\{\bigcup_{k\geq(1+\epsilon)k_0}(U_n(k)\geq U_n(k_0))\Big\} \leq \sum_{k\geq(1+\epsilon)k_0}^n P\{U_n(k)\geq U_n(k_0)\}
\leq c_1p^2n\left[e^{-c_2\tau}+ne^{-c_2\sqrt{\tau}}\right],
\end{eqnarray*}
 which follows if the estimate
\begin{equation}\label{yq0531.2}
P\{U_n(k)\geq U_n(k_0)\} \leq c_1p^2[e^{-c_2\tau}+ne^{-c_2\sqrt{\tau}} ]
\end{equation}
holds. By Assumption \ref{cond5} this term is of order $o\left(\frac{1}{n}\right)$
uniformly with respect to   $(1 + \varepsilon)k_0 \leq k \leq n$.
For a proof of this statement define the vectors $\bbw_i=(W_{i1},\cdots,W_{im})$ by
\[
\dot{\widetilde{\bbx}}_i=\bbw_i+\E\widetilde{\bbx}_i, 
\]
and denote by $\bbmu_1$ and $\bbmu_2$ the $m$-dimensional vectors containing the elements   of the matrices $\Sigma_1$ and $\Sigma_2$, respectively, which correspond to positions $(a,b) \in  \mathcal{D}_\tau$ identified in Step 1 of the procedure.
We will make use of the decomposition
\begin{eqnarray*}
U_n(k)&=&\frac{1}{n^4}\mathop{\sum^{k}\sum^{k}}_{(i\neq t)=1}\mathop{\sum^{n}\sum^{n}}_{(j\neq l)=k+1}
\left[(\bbw_i-\bbw_j)+(\E\widetilde{\bbx}_i-\E\widetilde{\bbx}_j)\right]^T
\left[(\bbw_t-\bbw_l)+(\E\widetilde{\bbx}_t-\E\widetilde{\bbx}_l)\right]\non
&=&A(k)+B(k)+C(k)+E(k),
\end{eqnarray*}
where
\begin{eqnarray*}
A(k)&=&\frac{1}{n^4}\mathop{\sum^{k}\sum^{k}}_{(i\neq t)=1}\mathop{\sum^{n}\sum^{n}}_{(j\neq l)=k+1}(\bbw_i-\bbw_j)^T(\bbw_t-\bbw_l), \\
B(k)&=&\frac{1}{n^4}\mathop{\sum^{k}\sum^{k}}_{(i\neq t)=1}\mathop{\sum^{n}\sum^{n}}_{(j\neq l)=k+1}(\E\widetilde{\bbx}_t-\E\widetilde{\bbx}_l)^T(\bbw_i-\bbw_j), \\
C(k)&=&\frac{1}{n^4}\mathop{\sum^{k}\sum^{k}}_{(i\neq t)=1}\mathop{\sum^{n}\sum^{n}}_{(j\neq l)=k+1}(\E\widetilde{\bbx}_i-\E\widetilde{\bbx}_j)^T(\bbw_t-\bbw_l), \\
E(k)&=&\frac{1}{n^4}\mathop{\sum^{k}\sum^{k}}_{(i\neq t)=1}\mathop{\sum^{n}\sum^{n}}_{(j\neq l)=k+1}(\E\widetilde{\bbx}_i-\E\widetilde{\bbx}_j)^T(\E\widetilde{\bbx}_t-\E\widetilde{\bbx}_l),
\end{eqnarray*}
and begin investigating the constant terms $E(k)$ and $E(k_0)$. For this purpose we make use of the notation
\[
a_{n,k}=  (n-k)(n-k-1)
\]
and obtain by a direct calculation
\begin{eqnarray}\label{yq0405.1}
&&E =E(k_0)-E(k)=\frac{1}{n^4}\big[k_0(k_0-1)a_{n,k_0}-k_0(k_0-1)a_{n,k}\big]\|\bbmu_1-\bbmu_2\|^2\non
&=&\frac{1}{n^4}k_0(k_0-1)\big[(n-k)(k-k_0)+(n-k-1)(k-k_0)+(k-k_0)^2\big]\|\bbmu_1-\bbmu_2\|^2. ~~~~
\end{eqnarray}
Observing the inclusion
\begin{equation}\label{H7}
\{ U_n(k) \geq U_n(k_0)\} \subset \{ A(k) + B(k) + C(k) -( A(k_0) + B(k_0) + C(k_0) )\geq E \}
\end{equation}
we now
  investigate the other terms $A(k) - A(k_0), B(k) - B(k_0), C(k) - C(k_0)$. A straightforward but tedious calculation yields the decomposition
\begin{eqnarray*}
B(k)-B(k_0)
= -B_1+B_2+B_3,
\end{eqnarray*}
where
\begin{eqnarray*}
&&B_1=\frac{1}{n^4}(k_0-1)[a_{n,k_0}-a_{n,k}](\bbmu_1-\bbmu_2)^T\sum_{i=1}^{k_0}\bbw_i,\non
&&B_2=\frac{1}{n^4}k_0(k_0-1)(k-k_0)(\bbmu_1-\bbmu_2)^T\sum_{j=k_0+1}^{n}\bbw_j,\non
&&B_3=\frac{1}{n^4}k_0(k_0-1)(n-k-1)(\bbmu_1-\bbmu_2)^T\sum_{j=k_0+1}^{k}\bbw_j.
\end{eqnarray*}
Using (\ref{yq0405.1})  the term $B_1$ can be handled as follows
\begin{eqnarray}\label{yq22.1}
\pr\{|B_1|\geq cE\} &=& \pr\Big\{\Big|(\bbmu_1-\bbmu_2)^T\sum_{i=1}^{k_0}\bbw_i\Big|\geq c k_0\|\bbmu_1-\bbmu_2\|^2\Big\}\non
&\leq&
\pr\Big\{
\big \| \sum_{i=1}^{k_0}\bbw_i \big \|^2 
\geq (c k_0\|\bbmu_1-\bbmu_2\|)^2  \Big\}\non
&\leq&
 \sum_{j=1}^m\pr\Big\{\Big|\sum_{i=1}^{k_0}W_{ij}\Big|\geq\tfrac{c k_0\|\bbmu_1-\bbmu_2\|}{\sqrt{m}}  \Big\}\non
&=&\sum_{j=1}^m\pr\Big\{\Big|\tfrac{1}{k_0}\sum_{i=1}^{k_0}W_{ij}\Big|\geq c\tfrac{\|\bbmu_1-\bbmu_2\|}{\sqrt{m}}  \Big\}.
\end{eqnarray}
Note  that for each $j=1,\cdots, m$, there exists a position $(a,b)$ such that $W_{ij}$ can be written as
\begin{equation}\label{yq22.2}
W_{ij}={\dot X}_{ia}{\dot X}_{ib}-\E(X_{ia}X_{ib})=X_{ia}X_{ib}-\E(X_{ia}X_{ib})-X_{ia}\bar X_b-X_{ib}\bar X_a+\bar X_a\bar X_b,
\end{equation}
and thus we obtain
\[
\frac{1}{k_0}\sum_{i=1}^{k_0}W_{ij}=\frac{1}{k_0}\sum_{i=1}^{k_0}\Big[X_{ia}X_{ib}-\E(X_{ia}X_{ib})\Big]-\bar X_{k_0a}\bar X_b-\bar X_{k_0b}\bar X_a+\bar X_a\bar X_b.
\]
With these notations the probability in (\ref{yq22.1}) can be further bounded using Lemmas \ref{lem2} and \ref{lem3}, that is
\begin{eqnarray}\label{yq11.7}
 \pr\{|B_1|\geq cE\}
&\leq& m\Big[\pr\Big\{\big|\tfrac{1}{k_0}\sum_{i=1}^{k_0}[X_{ia}X_{ib}-\E(X_{ia}X_{ib})]\big|\geq c_1\tfrac{\|\bbmu_1-\bbmu_2\|}{\sqrt{m}}  \Big\}+2\pr\Big\{\big|\bar X_{k_0a}\big|\geq c_2\sqrt{\tfrac{\|\bbmu_1-\bbmu_2\|}{\sqrt{m}}}  \Big\}\non
&&+3\pr\Big\{\big|\bar X_{b}\big|\geq c_3\sqrt{\tfrac{\|\bbmu_1-\bbmu_2\|}{\sqrt{m}}}  \Big\}+\pr\Big\{\big|\bar X_a\big|\geq c_4\sqrt{\tfrac{\|\bbmu_1-\bbmu_2\|}{\sqrt{m}} } \Big\}\Big]\non
&\leq&c_6m\Big[e^{-c_7\min\{k_0\frac{\|\bbmu_1-\bbmu_2\|^2}{m}, k_0\frac{\|\bbmu_1-\bbmu_2\|}{\sqrt{m}} \}}+
e^{-c_7k_0\frac{\|\bbmu_1-\bbmu_2\|}{\sqrt{m}}}+e^{-c_7n\frac{\|\bbmu_1-\bbmu_2\|}{\sqrt{m}}}\Big]\non
&\leq&c_6p^2\Big[e^{-c_7\min\{\tau,\sqrt{n\tau}\}}+e^{-c_7\sqrt{n\tau}}+e^{-c_7\sqrt{n\tau}}\Big]\leq c_8p^2e^{-c_7\min\{\tau,\sqrt{n\tau}\}},
\end{eqnarray}
where the last inequality follows from the fact
  that $k_0\frac{\|\bbmu_1-\bbmu_2\|^2}{m}\geq C\tau
$ and $k_0\frac{\|\bbmu_1-\bbmu_2\|}{\sqrt{m}}\geq C\sqrt{n\tau}$ if the smallest nonzero entry of $(\Sigma_1-\Sigma_2)$ satisfies $|\sigma_1({a,b})-\sigma_2({a,b}) |\geq \lambda \geq C\sqrt{\frac{\tau}{n}}\frac{n}{k_0}$. Note that Assumption \ref{cond5} gives $p^2n=o(e^{c\tau}), p^2n=o(e^{c\sqrt{n\tau}})$, and therefore it follows that
\begin{equation}\label{yq0531.4}
\pr\{|B_1|\geq cE\}=o\left(\frac{1}{n}\right).
\end{equation}
Similarly, each component of the vector $B_2$ can be represented as
\[
\frac{1}{n-k_0}\sum_{i=k_0+1}^{n}W_{ij}=\frac{1}{n-k_0}\sum_{i=k_0+1}^{n}\Big[X_{ia}X_{ib}-\E(X_{ia}X_{ib})\Big]-\bar Y_{k_0a}\bar X_b-\bar Y_{k_0b}\bar X_a+\bar X_a\bar X_b
\]
for some $(a,b) \in \mathcal{P}$.
We can find that
\begin{eqnarray*}
\pr\{|B_2|\geq cE\} &=&\pr\Big\{\Big|(\bbmu_1-\bbmu_2)^T\sum_{i=k_0+1}^{n}\bbw_i\Big|\geq c(n-k_0+n-k-1)\|\bbmu_1-\bbmu_2\|^2\Big\}\non
&\leq&\pr\Big\{\big  \|\bbmu_1-\bbmu_2\big \| \big \| \sum_{i=k_0+1}^{n}\bbw_i\big \| \geq c (n-k_0+n-k-1)\|\bbmu_1-\bbmu_2\|^2  \Big\}\non
&\leq&m\cdot\pr\Big\{\Big|\sum_{i=k_0+1}^{n}W_{ij}\Big|\geq\tfrac{c(n-k_0+n-k-1)\|\bbmu_1-\bbmu_2\|}{\sqrt{m}}  \Big\}\non
&\leq&m\cdot\pr\Big\{\Big|\frac{1}{n-k_0}\sum_{i=k_0+1}^{n}W_{ij}\Big|\geq\tfrac{c\|\bbmu_1-\bbmu_2\|}{\sqrt{m}}  \Big\}\non
&\leq& c_1m \Big[e^{-c_2\min\{(n-k_0)\frac{\|\bbmu_1-\bbmu_2\|^2}{m},(n-k_0)\frac{\|\bbmu_1-\bbmu_2\|}{\sqrt{m}} \}}+e^{-c_2(n-k_0)\frac{\|\bbmu_1-\bbmu_2\|}{\sqrt{m}}}+e^{-c_2n\frac{\|\bbmu_1-\bbmu_2\|}{\sqrt{m}}}\Big]\non
&\leq&  c_1p^2\Big[e^{-c_3\min\{\tau,\sqrt{n\tau}\}}+e^{-c_3\sqrt{n\tau}}+e^{-c_3\sqrt{n\tau}}\Big]\leq c_4p^2e^{-c_3\min\{\tau,\sqrt{n\tau}\}},
\end{eqnarray*}
where the last inequality follows from the fact that    $|\sigma_1({a,b})-\sigma_2({a,b}) |\geq C\sqrt{\frac{\tau}{n}}\frac{n}{n-k_0}$ for all $(a,b) \in \mathcal{P}$, which implies
$
(n-k_0)\frac{\|\bbmu_1-\bbmu_2\|^2}{m}\geq C\tau
$ and $(n-k_0)\frac{\|\bbmu_1-\bbmu_2\|}{\sqrt{m}}\geq C\sqrt{n\tau}$.
So we get
$$
\pr\{|B_2|\geq cE\}=o\left(\frac{1}{n}\right).
$$
In order to get a similar result for the term  $B_3$ we note that
\begin{eqnarray*}
\pr\{|B_3|\geq cE\}
&\leq& \mathbb{P} \Big \{\|\bbmu_1-\bbmu_2\|^2
{\sum_{j=1}^m(\sum_{i=k_0+1}^{k}W_{ij})^2}\geq c \Big[2(k-k_0)+\frac{(k-k_0)^2}{n-k-1}\Big]^2\|\bbmu_1-\bbmu_2\|^4  \Big\}\non
&\leq&m \mathbb{P} \Big  \{\Big|\sum_{i=k_0+1}^{k}W_{ij}\Big|\geq {c\Big[2(k-k_0)+\tfrac{(k-k_0)^2}{n-k-1}\Big]\|\bbmu_1-\bbmu_2\|} /\sqrt{m}  \Big \}\non
&\leq&m\cdot \mathbb{P} \Big
\{\Big|\tfrac{1}{k-k_0}\sum_{i=k_0+1}^{k}W_{ij}\Big|\geq\tfrac{2c\|\bbmu_1-\bbmu_2\|}{\sqrt{m}}  \Big \}\non
&\leq& c_1m \Big[e^{-c_2\min\{(k-k_0)\frac{\|\bbmu_1-\bbmu_2\|^2}{m},(k-k_0)\frac{\|\bbmu_1-\bbmu_2\|}{\sqrt{m}} \}}+e^{-c_2(k-k_0)\frac{\|\bbmu_1-\bbmu_2\|}{\sqrt{m}}}+e^{-c_2n\frac{\|\bbmu_1-\bbmu_2\|}{\sqrt{m}}}\Big]\non
&\leq& c_1m \Big[e^{-c_2\epsilon\min\{k_0\frac{\|\bbmu_1-\bbmu_2\|^2}{m},k_0\frac{\|\bbmu_1-\bbmu_2\|}{\sqrt{m}} \}}+e^{-c_2\epsilon k_0\frac{\|\bbmu_1-\bbmu_2\|}{\sqrt{m}}}+e^{-c_2n\frac{\|\bbmu_1-\bbmu_2\|}{\sqrt{m}}}\Big] \non
&\leq &
c_3p^2e^{-c_4\min\{\tau,\sqrt{n\tau}\}}.
\end{eqnarray*}
 By condition (\ref{yq0410.3}) in Assumption \ref{cond4} and Assumption \ref{cond5}, we get
\begin{equation}\label{yq0531.6}
\pr\{|B_3|\geq cE\}=o\left(\frac{1}{n}\right).
\end{equation}

Combining (\ref{yq0531.4})  and (\ref{yq0531.6}), we can conclude that
\begin{equation}\label{yq0410.1}
\pr\{|B(k)-B(k_0)|\geq cE\} \leq  c_1p^2e^{-c_2\min\{\tau,\sqrt{n\tau}\}}  =o\left(\frac{1}{n}\right)
\end{equation}
uniformly with respect to $k \geq k_0(1+ \varepsilon)$.
Similarly, one can see that the estimate
\begin{equation}\label{yq0410.2}
\pr\{|C(k)-C(k_0)|\geq cE\}  \leq  c_1p^2e^{-c_2\min\{\tau,\sqrt{n\tau}\}} =o\left(\frac{1}{n}\right).
\end{equation}
Finally, we investigate the terms $A(k)$ and $A(k_0)$ introducing the decomposition
\[
A(k)=A_1(k)-2A_2(k)+A_3(k),
\]
where
\begin{eqnarray*}
&&A_1(k)=\frac{1}{n^4}(n-k)(n-k-1)\mathop{\sum^{k}\sum^{k}}_{(i\neq t)=1}\bbw_i^T\bbw_t,\quad
A_3(k)=\frac{1}{n^4}k(k-1)\mathop{\sum^{n}\sum^{n}}_{(j\neq l)=k+1}\bbw_j^T\bbw_l,\non
&&A_2(k)=\frac{1}{n^4}(k-1)(n-k-1)\sum_{i=1}^k\sum_{l=k+1}^{n}\bbw_i^T\bbw_l.
\end{eqnarray*}
Then
\[
A(k)-A(k_0)=\big[A_1(k)-A_1(k_0)\big]-2\big[A_2(k)-A_2(k_0)\big]+\big[A_3(k)-A_3(k_0)\big].
\]
In order to get that $\pr\{|A(k)-A(k_0)|\geq cE\}=o\left(\frac{1}{n}\right)$, it is sufficient to show that $\pr\{|A_i(k)-A_i(k_0)|\geq cE\}=o\left(\frac{1}{n}\right)$, $i=1,2,3$.
In the following we only show that
$$\pr\{|A_1(k)-A_1(k_0)|\geq cE\} \leq c_1p^2[e^{-c_2\tau}+ne^{-c_2\sqrt{\tau}} ]
=o\left(\frac{1}{n}\right)$$ uniformly with respect to $k\geq (1 + \varepsilon)k_0$. The other two terms can be treated similarly.
Define
\begin{eqnarray*}
H_1=\frac{a_{n,k_0}-a_{n,k}}{n^4}\mathop{\sum^{k_0}\sum^{k_0}}_{i\neq t}\bbw_i^T\bbw_t,\quad
H_2=\frac{a_{n,k}}{n^4}\mathop{\sum^{k}\sum^{k}}_{i\neq t=(k_0+1)}\bbw_i^T\bbw_t,\quad H_3=\frac{a_{n,k}}{n^4}\sum_{i=1}^{k_0}\sum_{t=k_0+1}^k\bbw_i^T\bbw_t,
\end{eqnarray*}
and note that
\begin{eqnarray*}
A_1(k)-A_1(k_0)=-H_1+H_2+2H_3.
\end{eqnarray*}
First, we obtain for the term $H_1$ using (\ref{yq0410.3})
\begin{eqnarray}\label{yq11.6}
&&\pr\{|H_1|\geq cE\}=\pr\Big\{\Big|\mathop{\sum^{k_0}\sum^{k_0}}_{i\neq t}\bbw_i^T\bbw_t\Big|\geq c k_0(k_0-1) \|\bbmu_1-\bbmu_2\|^2 \Big\}\non
&=&\pr\Big\{\Big|\sum_{j=1}^m\mathop{\sum^{k_0}\sum^{k_0}}_{i\neq t}W_{ij}W_{tj}\Big|\geq c k_0(k_0-1) \|\bbmu_1-\bbmu_2\|^2 \Big\}\non
&\leq&m\cdot\pr\Big\{\Big|\mathop{\sum^{k_0}\sum^{k_0}}_{i\neq t}W_{ij}W_{tj}\Big|\geq c k_0(k_0-1) \tfrac{\|\bbmu_1-\bbmu_2\|^2}{m}  \Big\}\non
&\leq&m\cdot\pr\Big\{ \Big|\tfrac{1}{k_0}\sum_{i=1}^{k_0}W_{ij}\Big|\geq c_1 \tfrac{\|\bbmu_1-\bbmu_2\|}{\sqrt{m}}  \Big\}+m\cdot\pr\Big\{ \Big|\tfrac{1}{k_0}\sum_{i=1}^{k_0}W_{ij}^2\Big|\geq c(k_0-1) \tfrac{\|\bbmu_1-\bbmu_2\|^2}{m}  \Big\}\non
&\leq&  c_2p^2e^{-c_3\min\{\tau,\sqrt{n\tau}\}}+p^2\pr\Big\{ \Big|\tfrac{1}{k_0}\sum_{i=1}^{k_0}W_{ij}^2\Big|\geq c(k_0-1) \tfrac{\|\bbmu_1-\bbmu_2\|^2}{m}  \Big\},
\end{eqnarray}
where the last inequality follows by similar arguments as used in the derivation of   (\ref{yq22.1}) and (\ref{yq11.7}). For the second term, we use the decomposition (\ref{yq22.2})
\begin{eqnarray*}
&&p^2\pr\Big\{ \Big|\tfrac{1}{k_0}\sum_{i=1}^{k_0}W_{ij}^2\Big|\geq \tfrac{c}{2}(k_0-1) \tfrac{\|\bbmu_1-\bbmu_2\|^2}{m}  \Big\}\non
&\leq& p^2 k_0\cdot\pr\Big\{ \big|W_{ij}\big|\geq \sqrt{\tfrac{c}{2}(k_0-1)} \tfrac{\|\bbmu_1-\bbmu_2\|}{\sqrt{m}}  \Big\}\non
&\leq& p^2 k_0\cdot\pr\left\{ |X_{ia}X_{ib}-\E(X_{ia}X_{ib})|\geq c_4\sqrt{k_0} \tfrac{\|\bbmu_1-\bbmu_2\|}{\sqrt{m}}  \right\}\non
&&+2p^2 k_0\cdot\pr\left\{ |X_{ia}\bar X_b|\geq c_4\sqrt{k_0} \tfrac{\|\bbmu_1-\bbmu_2\|}{\sqrt{m}}  \right\}+p^2 k_0\cdot\pr\left\{ |\bar X_a\bar X_b|\geq c_4\sqrt{k_0} \tfrac{\|\bbmu_1-\bbmu_2\|}{\sqrt{m}}  \right\}\non
&\leq& c_5p^2 k_0e^{-c_6\sqrt{k_0}\frac{\|\bbmu_1-\bbmu_2\|}{\sqrt{m}}}
\leq c_5p^2 ne^{-c_7\sqrt{\tau}}=o\left(\tfrac{1}{n}\right),
\end{eqnarray*}
where the last line uses Lemma \ref{lem1} for the sub-exponential random variable $X_{ia}X_{ib}$ and the probability of this sub-exponential term is also the leading one among the remaining three terms. Moreover, because $|\sigma_1({a,b})-\sigma_2({a,b}) |\geq C\sqrt{\frac{\tau}{n}}\sqrt{\frac{n}{k_0}}$ for all $(a,b)\in \mathcal{P}$ we have
$\sqrt{k_0}\frac{\|\bbmu_1-\bbmu_2\|}{\sqrt{m}}\geq C\sqrt{\tau}$. Then the order $o\left(\frac{1}{n}\right)$  comes from the assumptions \eqref{c1} - \eqref{c3}.
Combining this estimate with \eqref{yq11.6} gives
$$
\mathbb{P} (|H_1| \geq cE) \leq c_2p^2e^{-c_3\min\{\tau,\sqrt{n\tau}\}}+ c_5p^2 ne^{-c_7\sqrt{\tau}}= o(\frac{1}{n}).
$$
Next for the term $H_2$, similarly  we can calculate
\begin{eqnarray}\label{yq11.8}
&&\pr\{|H_2|\geq cE\}\non
&=&\mathbb{P} \Big  \{\Big|\mathop{\sum^{k}\sum^{k}}_{i\neq t=(k_0+1)}\bbw_i^T\bbw_t\Big|\geq c \tfrac{k_0(k_0-1)}{n-k}\Big[\tfrac{n-k}{n-k-1}(k-k_0)+(k-k_0)+\tfrac{(k-k_0)^2}{n-k-1}\Big]\|\bbmu_1-\bbmu_2\|^2 \Big \}\non
&\leq&m\cdot \mathbb{P} \Big  \{\Big|\mathop{\sum^{k}\sum^{k}}_{i\neq t=(k_0+1)}W_{ij}W_{tj}\Big|\geq c \tfrac{k_0(k_0-1)}{n-k}\Big[2(k-k_0)+\tfrac{(k-k_0)^2}{n-k-1}\Big]\tfrac{\|\bbmu_1-\bbmu_2\|^2}{m} \Big \}\non
&\leq&
m\cdot  \mathbb{P} \Big  \{\Big|\frac{1}{k-k_0}\sum_{i=k_0+1}^k W_{ij}\Big|\geq \sqrt{\tfrac{c}{2} \tfrac{k_0(k_0-1)}{(n-k)}\Big[\tfrac{2}{(k-k_0)}+\tfrac{1}{n-k-1}\Big]}\tfrac{\|\bbmu_1-\bbmu_2\|}{\sqrt{m}} \Big \}\non
&&+m\cdot  \mathbb{P} \Big \{\Big|\tfrac{1}{k-k_0}\sum_{i=k_0+1}^k W_{ij}^2\Big|\geq \tfrac{c}{2} \tfrac{k_0(k_0-1)}{n-k}\Big[2+\tfrac{(k-k_0)}{n-k-1}\Big]\tfrac{\|\bbmu_1-\bbmu_2\|^2}{m} \Big \}.
\end{eqnarray}
Using similar arguments as in the discussion of the term (\ref{yq11.6}), the above probability can be further bounded by
\begin{eqnarray*}
&&\pr\{|H_2|\geq cE\}\non
&\leq& c_1m e^{-c_2 (k-k_0)\min\{\frac{k_0^2}{(n-k_0)(k-k_0)}\frac{\|\bbmu_1-\bbmu_2\|^2}{m},\sqrt{\frac{k_0^2}{(n-k_0)(k-k_0)}}\frac{\|\bbmu_1-\bbmu_2\|}{\sqrt{m}} \}}+c_2mne^{-c_3\sqrt{ \frac{k_0^2}{n-k_0}}\frac{\|\bbmu_1-\bbmu_2\|}{\sqrt{m}}}\non
&\leq&  c_1p^2e^{-c_4\min\{\tau,\sqrt{n\tau}\}}+c_2p^2ne^{-c_5\sqrt{\tau}}=o\left(\frac{1}{n}\right),
\end{eqnarray*}
where the last line is due to the observation that if the smallest nonzero entry of $(\Sigma_1-\Sigma_2)$ satisfies $|\sigma_1({a,b})-\sigma_2({a,b}) |\geq C\sqrt{\frac{\tau}{n}}\frac{n}{k_0}\sqrt{\frac{n-k_0}{k_0}}$ for some large $C$ in (\ref{yq0410.3}), then
$$
\frac{k_0^2}{(n-k_0)}\frac{\|\bbmu_1-\bbmu_2\|^2}{m}\geq C\tau
\mbox { and } \sqrt{\frac{k_0^2(k-k_0)}{n-k_0}}\frac{\|\bbmu_1-\bbmu_2\|}{\sqrt{m}}\geq C\sqrt{n\tau}.
$$
 So together with $p^2n=o(e^{c\tau}), p^2n=o(e^{c\sqrt{n\tau}})$  and $p^2n^2=o(e^{c\sqrt{\tau}})$ in Assumption \ref{cond5}, we can find the probability to be of order $o\left(\frac{1}{n}\right)$.
For the term $H_3$, according to (\ref{yq0410.3}) in Assumption \ref{cond4}, we have
\begin{eqnarray*}
&&\pr\{|H_3|\geq cE\}
\leq m\cdot \mathbb{P} \Big
\{\Big|\sum_{i=1}^{k_0}\sum_{t=k_0+1}^kW_{ij}W_{tj}\Big|\geq c \tfrac{k_0(k_0-1)}{n-k}\Big[2(k-k_0)\Big]\tfrac{\|\bbmu_1-\bbmu_2\|^2}{m} \Big \}\non
&\leq&m\cdot
\mathbb{P} \Big
\{\Big|\tfrac{\sum\nolimits_{i=1}^{k_0}W_{ij}}{k_0}\Big|\geq \sqrt{\tfrac{(k_0-1)}{n-k_0}}\tfrac{\|\bbmu_1-\bbmu_2\|}{\sqrt{m}} \Big \}
+m\cdot \mathbb{P} \Big \{\Big|\tfrac{\sum\nolimits_{t=k_0+1}^{k}W_{tj}}{k-k_0}\Big|\geq 2c\sqrt{\tfrac{(k_0-1)}{n-k_0}}\tfrac{\|\bbmu_1-\bbmu_2\|}{\sqrt{m}} \Big \}\non
&\leq&c_1m[e^{-c_2k_0\min\{\frac{k_0}{n-k_0}\frac{\|\bbmu_1-\bbmu_2\|^2}{m},\sqrt{\frac{k_0}{n-k_0}}\frac{\|\bbmu_1-\bbmu_2\|}{\sqrt{m}} \}}+ e^{-c_3(k-k_0)\min\{\frac{k_0}{n-k_0}\frac{\|\bbmu_1-\bbmu_2\|^2}{m},\sqrt{\frac{k_0}{n-k_0}}\frac{\|\bbmu_1-\bbmu_2\|}{\sqrt{m}} \}}]\non
&\leq&c_1me^{-c_4\min\{\tau,\sqrt{n\tau}\}}=o\left(\frac{1}{n}\right)
\end{eqnarray*}
when the smallest nonzero entry of $(\Sigma_1-\Sigma_2)$ satisfies $|\sigma_1({a,b})-\sigma_2({a,b}) |\geq C\sqrt{\frac{\tau}{n}}\frac{n}{k_0}\sqrt{\frac{n-k_0}{k_0}}$ for some large $C$ and $p^2n=o(e^{c\tau}), p^2n=o(e^{c\sqrt{n\tau}})$.
Combining these arguments gives
$$
\mathbb{P} (|A_\ell (k) - A_\ell(k_0)| \geq cE) \leq c_1p^2[e^{-c_2\tau}+ne^{-c_2\sqrt{\tau}} ]= o\big(\frac {1}{n}\big), \qquad \ell=1,2,3,
$$
where we note once again that the cases $\ell =2,3$ follow by similar arguments as given for $\ell =1$.
From \eqref{H7}, \eqref{yq0410.1}, \eqref{yq0410.2} we therefore obtain
\eqref{yq0531.2}, which proves
$$
\mathbb{P} \{ \hat k \geq (1+ \epsilon)k_0 \}  \leq c_5p^2n\left[e^{-c_6\tau}+ne^{-c_6\sqrt{\tau}}\right]
\rightarrow 0.
$$
By the discussion at the beginning of the proof and \eqref{yq0531.3}
the assertion of Theorem \ref{theorm2} follows.

\subsection{Proof of Corollary \ref{coro1}}
The difference  in proving Theorem \ref{theorem1}, \ref{theorem1a} and   \ref{theorm2} under Assumption \ref{cond5} and Assumption \ref{condnew} consists only in a different treatment of   the terms $C^{(2)}$ in (\ref{yq23}), $H_1$ in (\ref{yq11.6}) (and $H_2$ in (\ref{yq11.8})), for which we need to make use of the following Proposition \ref{prop1}. The proof of this result is postponed to Section \ref{prop}.

\begin{prop}\label{prop1}
Suppose $y_1,\cdots, y_k$ ($k\geq n^{\epsilon}$ for some $0<\epsilon<1$) are independent sub-exponential random variables.  Let $\Delta>\max\limits_{i}\E [y_i^2]$. Then
for any positive constants $c>0$, $M >0$ there exists a constant $n_0= n_0(c,M) \in \mathbb{N}$, such that for all $n \geq n_0$.
$$
\pr\Big\{\frac{1}{k}\sum_{i=1}^k y_i^2>\Delta\Big\}<cn^{-M}.
$$
\end{prop}
\noindent First, we discuss the differences in the proof of Theorem \ref{theorem1}   and look at the term $C^{(2)}$ in (\ref{yq23})
recalling the representation  $\dot{X}_{ia}\dot{X}_{ib}=X_{ia}X_{ib}-X_{ia}\bar X_{b}-X_{ib}\bar X_{a}+\bar X_{a}\bar X_b$. Proposition \ref{prop1} gives for the sum corresponding to the first term
\begin{eqnarray*}
\pr\Big\{\frac{1}{k}\sum_{i=1}^k({X}_{ia}{X}_{ib})^2>c\tau \Big\}\leq c\cdot n^{-M},\quad\forall M>0.
\end{eqnarray*}
Moreover, for $k=\lfloor n^{1/2} \rfloor +1,\ldots, n- \lfloor n^{1/2}\rfloor -1$ we have
\begin{eqnarray*}
 \pr\Big \{\tfrac{1}{k}\sum_{i=1}^k({X}_{ia}\bar{X}_{b})^2>c\tau \Big \}
 & \leq &  k\pr\left\{|{X}_{ia}\bar{X}_{b}|>c\sqrt{\tau} \right\}\non
&Ê\leq& n  \left[\pr\left\{|{X}_{ia}\bar X_b|>c\sqrt{\tau},|\bar X_b|>1\right\}+\pr\left\{|{X}_{ia}\bar X_b|>c\sqrt{\tau},|\bar X_b|\leq 1\right\}\right]\non
&Ê\leq &n  \left[\pr\left\{|\bar X_b|>1\right\}+\pr\left\{|{X}_{ia}|>c\sqrt{\tau}\right\}\right]\leq c_1n \left[e^{-c_2n}+e^{-c_3\tau}\right],\non
\pr\Big \{\tfrac{1}{k}\sum_{i=1}^k(\bar{X}_{a}\bar{X}_{b})^2>c\tau\Big \}
&\leq  & k\pr\left\{|\bar{X}_{a}\bar{X}_{b}|>c\sqrt{\tau} \right\}\non
&\leq & n \left[\pr\left\{|\bar{X}_{a}|>c_1\tau^{1/4} \right\}+\pr\left\{|\bar{X}_{b}|>c_2\tau^{1/4} \right\}\right]\leq c_3p^2n^2e^{-c_4n\sqrt{\tau}},
\end{eqnarray*}
and the probability in  \eqref{yq23}   can be bounded by
$$
 \pr\{C^{(2)}>c\tau\}\leq c_1\left[ n^{-M}+ n^2e^{-c_2n}+ n^2e^{-c_3\tau}+ n^2e^{-c_4n\sqrt{\tau}}\right].
$$
Consequently, if
$$
pn^{-M}\rightarrow 0, \quad p^2n^2e^{-cn}\rightarrow 0,\quad  p^2n^2 e^{-c\tau }\rightarrow 0
$$
for some small positive constant $c$, it follows that $p^2\cdot\pr\{C^{(2)}>c\tau\}\rightarrow 0$. Here $M>0$ could be any large positive constant. These estimates show
 that (\ref{yq0605.4}) holds for the case $i=2$ as long as
$$
pn^{-M}\rightarrow 0, \quad p^2n^2e^{-cn}\rightarrow 0,\quad  p^2n^2 e^{-c\tau }\rightarrow 0,\quad p^2n e^{-cn^{\frac{1}{4}}\sqrt{\tau}}\rightarrow 0.
$$
for some small positive constant $c$.
Note that  these  conditions contain  (\ref{yq0606.1})  and that
$pn^{-M}\rightarrow 0$ implies $p^2n^2e^{-cn}\rightarrow 0$ and $p^2n e^{-cn^{\frac{1}{4}}\sqrt{\tau}}\rightarrow 0$.
Consequently,   \eqref{yq001} holds  if
$$
pn^{-M}\rightarrow 0,\quad  p^2n^2 e^{-c\tau }\rightarrow 0,
$$
 where $c$ is some small positive constant and $M>0$ could be any large constant.

Next, we discuss the differences in the proof of Theorem \ref{theorem1a} and look exemplarily at the term $H_1$.
For the second term in (\ref{yq11.6}), recall that
 \[
 W_{ij}=X_{ia}X_{ib}-\E(X_{ia}X_{ib})-X_{ia}\bar X_b-X_{ib}\bar X_a+\bar X_a\bar X_b
 \]
 in equation (\ref{yq22.2}).  Proposition \ref{prop1} gives
 \begin{eqnarray*}
  \pr\Big\{ \Big|\tfrac{1}{k_0}\sum_{i=1}^{k_0}(X_{ia}X_{ib}-\E(X_{ia}X_{ib}))^2\Big|\geq \tfrac{c}{2}(k_0-1) \tfrac{\|\bbmu_1-\bbmu_2\|^2}{m}  \Big\}\leq c n^{-M}.
 \end{eqnarray*}
 In addition,
 \begin{eqnarray*}
 && \pr\Big\{ \Big|\tfrac{1}{k_0}\sum_{i=1}^{k_0}(X_{ia}\bar X_b)^2\Big|\geq \tfrac{c}{2}(k_0-1) \tfrac{\|\bbmu_1-\bbmu_2\|^2}{m}  \Big\}\leq   k_0\cdot\pr\left\{ |X_{ia}\bar X_b|\geq c_4\sqrt{k_0} \tfrac{\|\bbmu_1-\bbmu_2\|}{\sqrt{m}}  \right\}\non
 &\leq& n \left[\pr\left\{|{X}_{ia}\bar X_b|>c_4\sqrt{k_0} \tfrac{\|\bbmu_1-\bbmu_2\|}{\sqrt{m}},|\bar X_b|>1\right\}+\pr\left\{|{X}_{ia}\bar X_b|>c_4\sqrt{k_0} \tfrac{\|\bbmu_1-\bbmu_2\|}{\sqrt{m}},|\bar X_b|\leq 1\right\}\right]\non
&\leq& n \left[\pr\left\{|\bar X_b|>1\right\}+\pr\left\{|{X}_{ia}|>c_4\sqrt{k_0} \tfrac{\|\bbmu_1-\bbmu_2\|}{\sqrt{m}}\right\}\right]\leq c_1 n\left[e^{-c_2n}+e^{-c_3k_0\frac{\|\bbmu_1-\bbmu_2\|^2}{m}}\right],
 \end{eqnarray*}
and
\begin{eqnarray*}
  \pr\Big\{ \Big|\tfrac{1}{k_0}\sum_{i=1}^{k_0}(\bar X_{a}\bar X_b)^2\Big|\geq \tfrac{c}{2}(k_0-1) \tfrac{\|\bbmu_1-\bbmu_2\|^2}{m}  \Big\}&\leq& p^2 k_0\cdot\pr\left\{ |\bar X_{a}\bar X_b|\geq c_4\sqrt{k_0} \tfrac{\|\bbmu_1-\bbmu_2\|}{\sqrt{m}}  \right\}\non
&\leq&c_1p^2ne^{-c_2n\sqrt{k_0} \tfrac{\|\bbmu_1-\bbmu_2\|}{\sqrt{m}}  }.
\end{eqnarray*}
So, when the smallest nonzero entry of $(\Sigma_1-\Sigma_2)$ satisfies $|\sigma_1({a,b})-\sigma_2({a,b}) |\geq C\sqrt{\frac{\tau}{n}}\sqrt{\frac{n}{k_0}}$ for some large $C$,
$\sqrt{k_0}\frac{\|\bbmu_1-\bbmu_2\|}{\sqrt{m}}\geq C\sqrt{\tau}$, the second term can be bounded by
\begin{eqnarray*}\label{009.6}
\pr\Big \{ \Big|\tfrac{1}{k_0}\sum_{i=1}^{k_0}W_{ij}^2\Big|\geq \tfrac{c(k_0-1)}{2} \tfrac{\|\bbmu_1-\bbmu_2\|^2}{m}  \Big \}
&\leq& c[n^{-M}+ne^{-cn}+ne^{-ck_0\frac{\|\bbmu_1-\bbmu_2\|^2}{m}}+ne^{-cn\sqrt{k_0} \frac{\|\bbmu_1-\bbmu_2\|}{\sqrt{m}}  }]\non
&\leq&c[n^{-M}+ne^{-cn}+ne^{-c\tau}+ ne^{-cn\sqrt{\tau}}],
\end{eqnarray*}
and we obtain
\[
\pr\{|H_1|\geq cE\}\leq  c[p^2e^{-c\min\{\tau,\sqrt{n\tau}\}}+p^2n^{-M}+p^2ne^{-cn}+p^2ne^{-c\tau}+ p^2ne^{-cn\sqrt{\tau}}]=o\left(\frac{1}{n}\right),
\]
where  the last estimate follows from  $pn^{-M}\rightarrow 0$ and  $pn e^{-c\tau }\rightarrow 0$ in Assumption \ref{condnew}.

Finally,  the term $H_2$ in (\ref{yq11.8})
can be treated using the same derivation as in (\ref{009.6}) and  we obtain
\begin{eqnarray*}
 \pr\{|H_2|\geq cE\}\leq  c_1p^2e^{-c_4\min\{\tau,\sqrt{n\tau}\}}+c_2p^2[n^{-M}+ne^{-cn}+ne^{-c\tau}+ ne^{-cn\sqrt{\tau}}]=o\left(\frac{1}{n}\right),
\end{eqnarray*}
where we use the fact  that
$$
\frac{k_0^2}{(n-k_0)}\frac{\|\bbmu_1-\bbmu_2\|^2}{m}\geq C\tau
$$
if the smallest nonzero entry of $(\Sigma_1-\Sigma_2)$ satisfies $|\sigma_1({a,b})-\sigma_2({a,b}) |\geq C\sqrt{\frac{\tau}{n}}\frac{n}{k_0}\sqrt{\frac{n-k_0}{k_0}}$ for some large $C$ in (\ref{yq0410.3}).
Together with  the conditions $pn^{-M}\rightarrow 0$ and  $pn e^{-c\tau }\rightarrow 0$  from Assumption \ref{condnew} it follows that  the probability is
of order  $o\left(\frac{1}{n}\right)$.

Adjusting the above three terms in the proof of Theorem \ref{theorem1}, Theorem \ref{theorem1a} and Theorem \ref{theorm2}, we complete the proof of Corollary \ref{coro1}.

\subsection{Proof of Proposition \ref{prop1}}\label{prop}
Denote $C\triangleq \Delta-\frac{1}{k}\sum\limits_{i=1}^k \E y_i^2$, $C>0$. By   Theorem 4.1 in \cite{John} and the following remark, we have for any $h>2$
\begin{eqnarray}\label{09.2}
 \pr\Big\{\tfrac{1}{k}\sum_{i=1}^k y_i^2>\Delta\Big\}&=&\pr\Big\{\tfrac{1}{k}\sum_{i=1}^k (y_i^2-\E y_i^2)>C\Big\}\leq \tfrac{1}{C^hk^h}\E\Big[\sum_{i=1}^k (y_i^2-\E y_i^2)\Big]^h\\ \nonumber
&\leq&\tfrac{C_1^h}{C^hk^h}\Big(\tfrac{h}{\log h}\Big)^h\cdot
\max\Big\{\Big[\E\Big|\sum_{i=1}^k (y_i^2-\E y_i^2)\Big|^2\Big]^{\frac{h}{2}},\ \sum_{i=1}^k\E|y_i^2-\E y_i^2|^h\Big\},
\end{eqnarray}
where $C_1$ is an absolute constant. We calculate the upper bounds for the two terms using the fact that the random variables $y_i$  are sub-exponential. For $i=1,\ldots,k$ we have
\begin{eqnarray*}
\E|y_i^2-\E y_i^2|^h
&\leq& 2^{h+1} \E y_i^{2h}
=2^{h+1} [\E y_i^{2h}I\{|y_i|\leq h(\log h)^2\}+\E y_i^{2h}I\{|y_i|>h(\log h)^2\}]\non
&\leq&2^{h+1}[h^{2h}(\log h)^{4h}+c_1]\leq c_2\cdot 2^h h^{2h}(\log h)^{4h},
\end{eqnarray*}
and thus for any positive integer $M>0$, we can find $h>2$ such that
\begin{eqnarray}\label{09.3}
\frac{C_1^h}{C^hk^h}\Big(\frac{h}{\log h}\Big)^h\cdot\sum_{i=1}^k\E|y_i^2-\E y_i^2|^h\leq c_3^h k\cdot\Big[\frac{(h\log h)^3}{k}\Big]^h\leq cn^{-M},
\end{eqnarray}
where the last inequality is due to the fact that $k>n^{\epsilon}$.
Moreover, since the  random variables $y_i$ are independent, we obtain
\begin{eqnarray*}
 \Big[\E|\sum_{i=1}^k (y_i^2-\E y_i^2)|^2\Big]^{\frac{h}{2}}=\Big[\sum_{i=1}^k\E(y_i^2-\E y_i^2)^2\Big]^{\frac{h}{2}}\leq c_4^h k^{\frac{h}{2}}.
\end{eqnarray*}
Therefore, for any positive integer $M>0$, there exists a constant $h$, such that
\begin{eqnarray}\label{09.4}
\frac{C_1^h}{C^hk^h}\Big(\frac{h}{\log h}\Big)^h\cdot\Big[\E|\sum_{i=1}^k (y_i^2-\E y_i^2)|^2\Big]^{\frac{h}{2}}\leq c_5^{h}\Big(\frac{h}{\sqrt{k}\log h}\Big)^h\leq cn^{-M},
\end{eqnarray}
where the last inequality is also based on the fact that $k>n^{\epsilon}$. Combining (\ref{09.2}), (\ref{09.3}) and (\ref{09.4}) the assertion of Proposition \ref{prop1} follows.

\end{document}